\documentclass[12pt]{article}

\usepackage{setspace,graphicx,epstopdf,amsmath,amsfonts,amssymb,amsthm,versionPO}
\usepackage{marginnote,datetime,enumitem,subfigure,rotating,fancyvrb}
\usepackage{hyperref,float}
\usepackage{natbib}

\usepackage{algorithm}
\usepackage[noend]{algpseudocode}
\usepackage{appendix}
\usepackage{auto-pst-pdf}
\usepackage{booktabs}
\usepackage{cases}
\usepackage{dcolumn}
\usepackage{epsfig}
\usepackage{epigraph}
\usepackage{longtable}
\usepackage{lscape}
\usepackage[cal=boondox]{mathalfa}
\usepackage{mathpazo}
\usepackage{multirow}
\usepackage{stmaryrd}
\usepackage{threeparttable}
\usepackage{psfrag}
\usepackage{pst-node,pst-tree}
\usepackage{tabularx}

\usdate
\newcolumntype{.}[1]{D{.}{.}{#1}}

\excludeversion{notes}      
\includeversion{links}          

\iflinks{}{\hypersetup{draft=true}}

\ifnotes{%
\usepackage[margin=1in,paperwidth=10in,right=2.5in]{geometry}%
\usepackage[textwidth=1.4in,shadow,colorinlistoftodos]{todonotes}%
}{%
\usepackage[margin=1in]{geometry}%
\usepackage[disable]{todonotes}%
}

\makeatletter\let\chapter\@undefined\makeatother 

\usepackage{indentfirst} 
\usepackage{jfe}          

\begin{document}

\setlist{noitemsep}  

\title{Evaluating Conditional Cash Transfer Policies with Machine Learning Methods}

\author{Tzai-Shuen Chen\footnote{
I am deeply indebted to my advisor Juan Pantano and Werner Ploberger for their guidance throughout this project.
All errors are mine. Correspondence:chen.tzai-shuen\@wustl.edu.}\\
Washington University in St. Louis}
\date{}              


\renewcommand{\thefootnote}{\fnsymbol{footnote}}

\singlespacing

\maketitle

\vspace{-.2in}
This paper presents an out-of-sample prediction comparison between major machine learning models and the structural econometric model. Over the past decade, machine learning has established itself as a powerful tool in many prediction applications, but this approach is still not widely adopted in empirical economic studies. To evaluate the benefits of this approach, I use the most common machine learning algorithms, CART, C4.5, LASSO, random forest, and adaboost, to construct prediction models for a cash transfer experiment conducted by the Progresa program in Mexico, and I compare the prediction results with those of a previous structural econometric study. Two prediction tasks are performed in this paper: the out-of-sample forecast and the long-term within-sample simulation. For the out-of-sample forecast, both the mean absolute error and the root mean square error of the school attendance rates found by all machine learning models are smaller than those found by the structural model. Random forest and adaboost have the highest accuracy for the individual outcomes of all subgroups. For the long-term within-sample simulation, the structural model has better performance than do all of the machine learning models. The poor within-sample fitness of the machine learning model results from the inaccuracy of the income and pregnancy prediction models. The result shows that the machine learning model performs better than does the structural model when there are many data to learn; however, when the data are limited, the structural model offers a more sensible prediction. In addition to prediction outcome, machine learning models are more time-efficient than the structural model. The most complicated model, random forest, takes less than half an hour to build and less than one minute to predict. The findings of this paper show promise for adopting machine learning in economic policy analyses in the era of big data.


\medskip

\noindent \textit{JEL classification}: C18, I25, O15, O22.

\medskip
\noindent \textit{Keywords}: Cash transfers, Machine learning, Random forest, Out-of-sample prediction.

\thispagestyle{empty}

\clearpage

\onehalfspacing
\setcounter{footnote}{0}
\renewcommand{\thefootnote}{\arabic{footnote}}
\setcounter{page}{1}

\section{Introduction}
\noindent

Machine learning has regained its popularity in many research areas over the past decade due to
the availability of large data sets and the rapid growth in computing power. The advantages 
of machine learning are that it can efficiently extract information from huge volumes of data 
and build prediction models with excellent out-of-sample forecasts, which are usually the main concerns 
of policy makers and voters alike. 
This approach, however, is still not widely adopted in empirical economic studies. 
There are two reasons that might prohibit economists from embracing machine learning methods. 
The first reason is the Lucas Critique. A policy intervention would affect the incentive that people face and would thus change the underlying decision-making problem. A predictive model based on machine learning usually lacks 
a structural description of the optimizing behavior and might thus not capture people's reactions to
the policy intervention to give a reliable prediction. 
Second, even if machine learning models could tackle the previous problem, 
they might have no advantage over a well-performed standard econometric model 
in terms of prediction or time efficiency; hence, there is no need for economists to adopt the new methods. 

Previous studies have revealed the predictive power of machine learning in several economic applications.
\citet{Ahmed2010} compare different machine learning models for time series forecasting problems. 
\citet{Yang2016} use data from Zillow to evaluate the impact of fracking on nearby housing prices.
\citet{Bajari2015a} use several machine learning models to estimate the demand for salted snacks and 
compare the out-of-sample prediction with reduced form econometric model. 
\citet{Kleinberg2017} show that machine learning can be used to improve a judge's bail decisions.
However, no study has addressed the concern raised by the Lucas critique or compared the 
machine learning method with the structural econometric model. 
In this paper, I use the famous randomized conditional cash transfer (CCT) experiment 
conducted by the Progresa\footnote{The program was renamed Oportunidades in 2000 and Prospera in 2014} program in Mexico to investigate the answers to the previous two questions and 
explore the possibility of adopting machine learning in empirical economic research.

Conditional cash transfer has become a popular approach in developing countries to promote human capital investment, such as education, nutrition, and health-related activities, among poor families. People who have access to such a policy receive a certain amount of cash benefits only if certain requirements --- sending a child to school, receiving immunization, or visiting the health center on regular basis --- are satisfied. The Progresa program is one of the first institutes to offer the CCT to vulnerable households. To evaluate the effect of the conditional cash transfer, the program implemented a large-scale randomized experiment from 1997-1999 in the rural communities of Mexico, the results of which led to the growth of CCT policy in many other countries. Although previous studies have demonstrated the effectiveness of this or similar CCT policies \citep{Schultz2004, Gertler2004, Todd2006,Ranganathan2012}, most such studies focus on providing a causal explanation rather than predicting the policy outcomes. Although a good explanatory model might also be able to offer good predictive power, in most cases, its prediction performance rarely matches that of a predictive model \citet{Shmueli2010}.  

\citet{Todd2006} (henceforth, TW2006) use the Progresa experiment to evaluate the efficacy of their structural model for 
policy suggestion. Their work is one of the few studies in the economic literature to perform out-of-sample forecasting to test a model's predictive power while estimating the model coefficients. In their study, children are divided into eight subgroups by age, gender, and school performance for each year and treatment. The prediction difficulty increases both as the children's age increases because they have choices other than attending school and as the restriction of school performance increases due to the smaller sample sizes for learning. The out-of-sample predictions of the structural model are very close to the real outcomes in terms of school attendance rates across different subgroups. For most of the subgroups, the prediction errors are within two percentage points. The errors are approximately eight percentage points for the last two subgroups, boys and girls who are at least 13 years old, are behind by one or more years in school, and have completed primary education. The outstanding results of TW2006 offer an ideal benchmark for comparing the structural econometric models with machine learning models, which would reveal how much benefit an economist could derive from the new approach, and the Progresa randomized experiment provides us with a good opportunity to test the quantitative effect of the Lucas critique for CCT.

This paper tests five machine learning algorithms based on their general prediction performance and popularity, CART, C4.5, LASSO, Random Forest, and Adaboost\footnote{Although deep neural network (DNN) has achieved many successes in areas such as image recognition and artificial intelligence, this study does not test this approach because a good DNN relies on extremely large data sets and numerous high-performance GPUs. Given the data size of this study, the prediction outcome of DNN would not differ significantly from that of another algorithm.}, on two tasks.
The first task is the out-of-sample forecast, called one-period-ahead prediction by TW2006, which is the main focus for selecting prediction models. The second task is long-term within-sample simulation, called $N$-period-ahead prediction, which would help the researcher investigate more details of the model's properties between structural and machine learning approaches.

For the out-of-sample forecast, all the predictions of the machine learning models match the performance of the structural model in terms of both the mean absolute error (MAE) and root mean square error (RMSE). The ranking indicated by the MAE of the attendance rates, from best to worst, is Adaboost, random forest, LASSO, CART, C4.5, and structural model. When ranking by RMSE, again from best to worst, we get Adaboost, LASSO, random forest, CART, C4.5, and the structural model. Machine learning models tend to perform better than the structural model in the first and last two subgroups, but they perform worse in the third through sixth subgroups. The improvements in the first two subgroups are relatively small because the structural model already has close predictions. However, prediction errors of most machine learning models are decreased by at least 4 percentage points in the last category. The largest prediction error of all the machine learning models across all subgroups is 6.7 percentage points, while that of TW2006 is 8.4 points. 

In addition to the attendance rate comparison, I calculate the accuracy of each machine learning model to measure their performance from another perspective. The prediction accuracy is defined as follows:
\[
   \mbox	{Acc }= \frac{tp + tn}{ tp + fp + fn + tn},
\]
where $tp$ is the number of true positives, $tn$ is the number of true negatives, $fp$ is the number of false positives, 
and $fn$ is the number of false negatives. In other words, accuracy is the percentage of correct predictions. The accuracy measurement can give us a better understanding of the prediction performance for binary outcomes since the false positive and false negative rates can cancel each other out and still result in a good prediction of the average value. Table \ref{tab:AccExample} presents an example with perfect attendance rate prediction but zero accuracy. The actual attendance rate in this example is 50 percent for 100 students, and the predicted attendance is exactly 50 percent. However, the prediction for each individual student, as the confusion matrix shows, is never correct. In this case, we may prefer some models with small prediction errors but higher accuracy to the zero prediction error and zero accuracy model if the policy maker wants to maximize the cash transfer efficiency. 
That is, by using a high-accuracy model, we could identify individuals who would or would not be affected by the target policy 
and could hence offer a localized or even individualized solution, as suggested by \citet{DeJanvry2006}\footnote{\citet{DeJanvry2006} shows that 36 percent of dropouts do not continue school because of non-financial reasons}.

\begin{table}[htbp]
  \centering
  \caption{An Example with Perfect Attendance Rate Prediction but Zero Accuracy}
    \begin{tabular}{rrrr}
    \toprule
     Model  &       &\multicolumn{2}{c}{Actual Outcome}  \\
   \cmidrule(l){3-4}
               &            & Attend & Not Attend \\
    \midrule
    Prediction & Attend     &  0       & 50 \\
               & Not Attend & 50       &  0 \\
    \bottomrule
    \end{tabular}\hspace{.05\textwidth}
  \label{tab:AccExample}
\end{table}

The ranking of the machine learning models in terms of accuracy is random forest, adaboost, CART, LASSO, and C4.5.
For the first two subgroups, all models have close to 100 percent accuracy, which results from the fact that most children 
between 6 and 11 years of age do go to primary school. The accuracy decreases as the age and the school performance restriction increase. The best model, random forest, has accuracies of approximately 85 percent for the third to sixth subgroups and of approximately 80 percent for the last two. In contrast, the worst model, C4.5, has only 75 percent accuracy for the middle subgroups and 67 percent accuracy for the seventh subgroup. 


For the long-term within-sample simulation, the structural model performs better than do all of the machine learning models. The prediction errors of the long-term simulation are larger than those of the out-of-sample forecast for all models because the errors accumulate over time. The structural model manages to restrict the error to within 10 percentage points across all subgroups, while some of the machine learning models can have errors as large as 30 percentage points. There are several reasons that might cause the large errors observed for machine learning models in the long-term simulation: the inaccuracy of the income, pregnancy, and pregnancy prediction models. In a long-term simulation, the income model can only use a few variables that are available when a couple gets married to predict the household income, which leads to a homogeneous income prediction and large root mean square errors. The magnitude of the root mean square error can be several times the size of the subsidy. Since previous studies have shown that subsidies, which are part of the household income, have significant effects on school attendance decisions, the large errors in predicted income would also have huge impacts on the quality of the simulation.

The inaccuracy of the pregnancy model comes from both inaccurate income prediction and imbalanced data. The pregnancy rate in the training data is 12.94 percent, i.e., 489 of the 3780 training households. Under these circumstances, the machine learning model tends to under-predict the chance of pregnancy because a model predicting no pregnancy would still have an accuracy of 87.06 percent. The school failure prediction model suffers from a similar problem. The students who fail their current grade account for approximately 10 percent of the children who go to school. Since the number of years by which a child is behind in school also plays a key role in school attendance decisions, the inaccuracy of school failure could also be another reason for the large simulation error. This study employs several strategies to solve the imbalanced data problem, including undersampling, oversampling, and synthetic minority over-sampling (SMOTE). However, the best model still cannot fully overcome the problem. The result of long-term within-sample simulations reveal that machine learning models depend heavily on the quality, sample size, and covariate size of the data but that the structural econometric model can overcome these problems by imposing choice constraints.

In summary, the results of this study show that the extent to which the policy intervention undermines the validity of a prediction model, if it exists, is very mild for the conditional cash transfer policy.
Therefore, although machine learning algorithms do not contain any behavioral description,
they can still catch important interactions between the covariates and to offer accurate out-of-sample forecasts.
The prediction performance of the machine learning models would be better than that of the structural model if there were sufficient data for learning. However, when the data are limited, the structural model tends to offer more sensible predictions. 
In addition, machine learning models are more time-efficient than the standard structural model is. 
In this study, the most time-consuming model, random forest, takes approximately 30 minutes to build on both the Matlab and R platforms without any parallel computing. Given the increasing size of data sets, it is very likely that the out-of-sample forecast and time-efficient advantages of machine learning models will be increasingly apparent. The findings in this paper show the great potential of applying machine learning to future economic policy analyses.


The rest of the paper is organized as follows. Section 2 outlines the algorithms for the five machine learning models.
Section 3 describes the data and background of the Progresa program. Section 4 outlines the model training process for the school attendance, income prediction, pregnancy, and school failure models.
Section 5 presents the result of the out-of-sample prediction and long-term within-sample simulation.
Section 6 presents our discussion and conclusions.

\section{Machine learning algorithms} \label{sec:Model}
\subsection{Classification and regression tree (CART)}
\begin{figure}
\minipage{0.43\textwidth}
	\psfrag{a}{$\textbf{X}_1$} \psfrag{b}{$\textbf{X}_2$}
	\psfrag{c}{$v_{1,1}$}\psfrag{d}{$v_{2,2}$}\psfrag{e}{$v_{2,1}$}
	\psfrag{f}{$S_1$}\psfrag{g}{$S_2$}\psfrag{h}{$S_4$}\psfrag{I}{$S_3$}
	\includegraphics[width=0.9\linewidth]{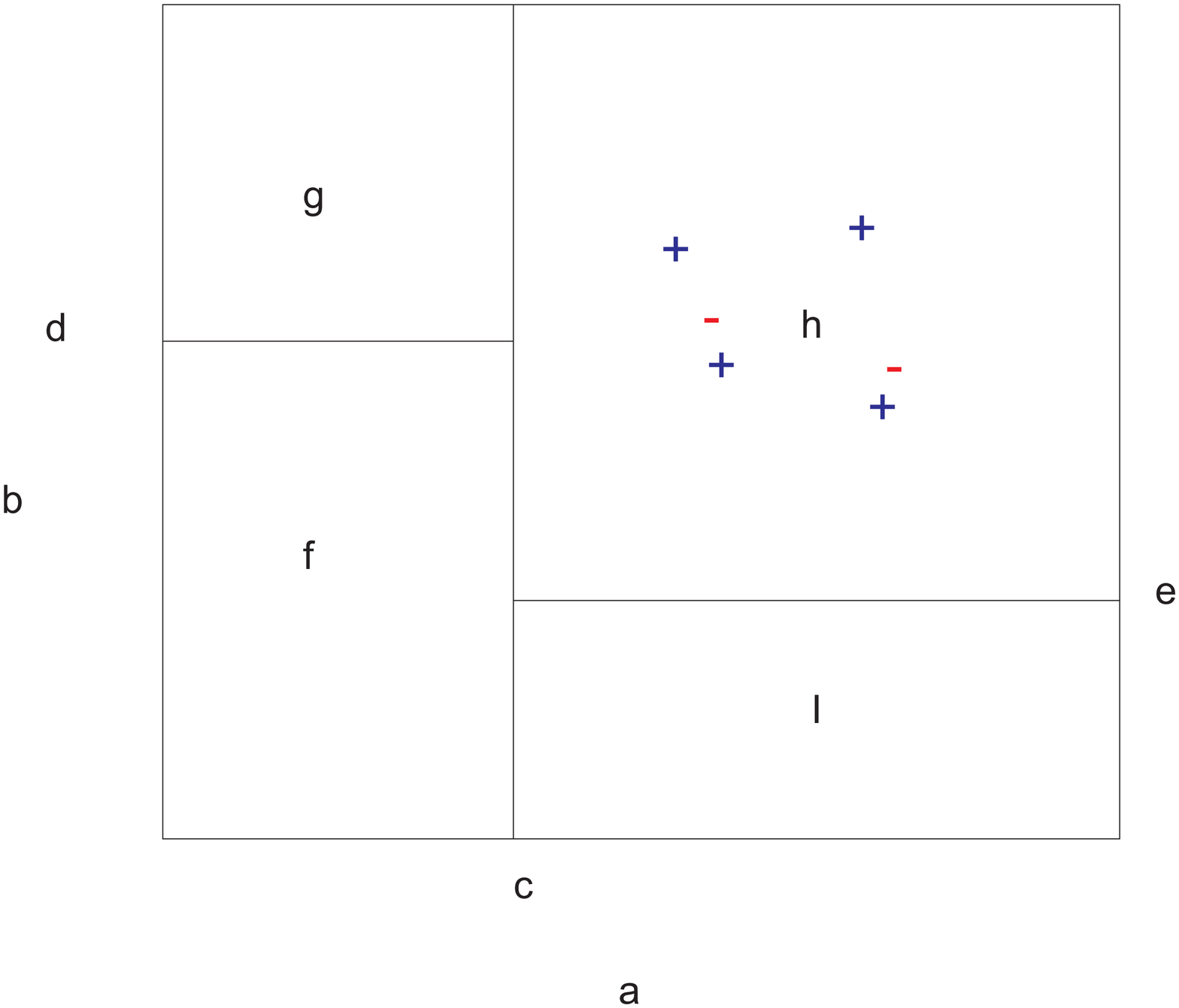}
\endminipage\hfill
\minipage{0.48\textwidth}
	\psset{labelsep=3pt,tnpos=a,radius=2pt}
	\pstree[treesep=65pt]{\TC*~{$X_1$}}{
		\pstree[levelsep=80pt]{\TC*~{}~[tnpos=l]{$X_2$}\tlput{$X_1 \leq v_{1,1}$}}{
			\TC*~{}~[tnpos=b]{$S_1$}\tlput{$X_2 \leq v_{2,2}$}		
			\TC*~{}~[tnpos=b]{$S_2$}\trput{$X_2 > v_{2,2}$}
		}			
		\pstree[levelsep=80pt]{\TC*~{}~[tnpos=r]{$X_2$}\trput{$X_1 > v_{1,1}$}}{
			\TC*~{}~[tnpos=b]{$S_3$}\tlput{$X_2 \leq v_{2,1}$}
			\TC*~{}~[tnpos=b]{$S_4$}\trput{$X_2 > v_{2,1}$}
		}
	}
\endminipage
\caption{CART}\label{fig:CART}
\end{figure}

CART is one of the most basic machine learning algorithms. The idea of this algorithm is to sequentially divide the input space 
$\textbf{X}_{N\times K}=(\textbf{X}_1, \textbf{X}_2, \cdots, \textbf{X}_K)$ into many subspaces, $S_i \in S$, 
and then use the average or mode of the responses in a subspace as the prediction of new data if its covariates satisfy the restrictions of this subspace. 
For example, consider a prediction problem with the binary outcome $Y$ and two-dimensional inputs $X_1$ and $X_2$.
Suppose that the covariates $(X_1, X_2)$ fall into subspace $S_4$, as shown in the left panel of Figure \ref{fig:CART}, i.e., 
$X_1 > v_{1,1}$ and $X_2 > v_{2,1}$, then the prediction of $Y$ would be $+1$, which is the mode of response in $S_4$.
This partition process can be represented as a binary tree model, as shown in the right panel of Figure \ref{fig:CART}.
The process usually employs a greedy approach to divide the input space since finding the optimal partition is an NP-complete problem \citep{Hyafil1976}. Let $V_k$ denote the set of realized values of $X_k$ in the training data. 
Then, at each tree node, the split function chooses a covariate $X_k$ and a threshold $v_{k,j} \in V_k$ to solve
\begin{align}	
	\max_{X_k, v_{k, j}} cost(D) - (\frac{|D_L|}{|D|}cost(D_L) + \frac{|D_R|}{|D|}cost(D_R),
\end{align}
where 
\[
	D=\{(X,Y)\} \mbox{, }D_L =\{(X,Y): X_k \leq v_{k,j}\} \mbox{, and } D_R =\{(X,Y): X_k > v_{k,j}\}.
\]
For a regression problem, the common choice for the cost function is the square loss,
\[
	cost(X,Y) = \sum (Y - avg(Y))^2.
\]

If the response $Y$ is a category variable, then define $p_{c,D}$ as 
\[p_{c,D} = \frac{1}{|D|}\sum_{Y \in D}I(Y=c)\],
and either Gini impurity or entropy can be used as the cost function.
\footnote{More specifically, CART, as developed by \citet{Breiman1984}, uses Gini impurity as the cost function, 
and the ID3 developed by \citet{Quinlan1979} uses entropy as the cost function. 
Since the concepts of both algorithms are very similar and their prediction performances usually do not differ significantly,
I refer to both methods as CART here.}
\[
	\mbox{Gini impurity: } \sum_{c=1}^{C} p_{c, D}(1-p_{c, D}),
\]
\[
	\mbox{Entropy: } -\sum_{c=1}^{C} p_{c, D}\log(p_{c, D}).
\]

CART is a primitive ML model and usually does not make as accurate predictions as do other more sophisticated ML models. 
Its prediction performance can therefore be treated as a lower bound for machine learning models.

\subsection{C4.5}
C4.5 is an improved version of CART that corrects the problem that CART tends to favor covariates with large numbers of values when using entropy as the cost function. Suppose that covariate $X_k$ has $m$ thresholds of $v_{k,j}$, and denote $D_{kj}=\{(X,Y): X_k=v_{k,j}\}$. Then, the cost function in C4.5 can be defined as a normalized information gain as follows:
\[
	\mbox{Normalized information gain: } \frac{Gain(X_k, v_{k,j})}{SplitInfo(X_k, v_{k,j})},
\]
where
\[
	Gain(X_k, v_{k,j}) = -\sum_{p_{c, D}} p_{c, D}\log(p_{c, D}) + (\frac{|D_L|}{|D|}\sum_{p_{c, D_L}} p_{c, D_L}\log(p_{c, D_L}) + \frac{|D_R|}{|D|} p_{c,D_R}\log(p_{c,D_R}),
\]
and
\[
	SplitInfo(X_k, v_{k,j}) = -\sum_{j=1}^{m} \frac{|D_{kj}|}{|D|}\log(\frac{|D_{kj}|}{|D|}).
\]

\subsection{LASSO}
LASSO is a regression model with $L_1$ regularization:
\[
	\min ||y-X\beta||_{2}^{2} +\lambda ||\beta||_{1}.
\] 
The $L_1$ norm gives LASSO the ability of variable selection, which is useful for increasing the prediction accuracy when there is a large number of covariates in the data. LASSO can also be viewed as a special case of the support vector machine (SVM) approach. The idea of SVM is to find a boundary such that the data points in each subgroup are as far from the boundary as possible.

\subsection{Random Forest}
Random Forest is an ensemble machine learning algorithm that uses a bootstrapping method to reduce the variance of prediction. This method first applies sampling with replacement to produce $T$ samples, each with the same data size as the original data, and then uses each sample to build a decision tree, $f_t(X)$. The prediction of new data is then the average or mode of the predictions from each decision tree. Hence, the prediction is the outcome of a ``forest.''
\[
	\hat{f}(X^{'}) = \frac{1}{T} \sum_{1}^{T} f_t(X) \mbox{ (regression) or \textit{majority votes of }} \{f_t(X)\}_{1}^{T} 
	\mbox{ (classification)}
\]
The ``random'' part of the name comes from the fact that a procedure randomly selects $p$ covariates out of $\textbf{X}=(\textbf{X}_1, \textbf{X}_2, \cdots, \textbf{X}_K)$ to determine the splitting covariate and value at each node. This step can reduce the correlation of each tree in the forest and, hence, reduce the variance of the prediction. Algorithm \ref{rf} shows the pseudocode of random forest.

\begin{algorithm}
\caption{Random Forest}\label{rf}
\begin{algorithmic}[1]
\For{$t = 1 \to T$} 
\State Draw a sample of size $N$ from the training data, $X_{N\times K}$
\Procedure{GrowTree}{$X$}
	\While{Node size $>$ minimum node size and the node is not pure}
		\State{Select $p$ covariates out of $i$ variables}
		\State{Decide the splitting covariate and splitting value from the $p$ covariates}
		\State{Divide current data into two groups, $X_{L}$ and $X_{R}$.}
		\State{\Call{GrowTree}{$X_{L}$}}
		\State{\Call{GrowTree}{$X_{R}$}}
		\State{\Return}
	\EndWhile
\EndProcedure
\EndFor
\State Average all trees
\end{algorithmic}
\end{algorithm}

\subsection{Adaboost}
\citet{Schapire1990} showed that the performance of a weak learner, defined as any algorithm that only performs slightly better than random guessing, can be boosted by adaptively combining learners. A boosting algorithm tries to minimize a loss function with the
following form of the model at the $s$th iteration:
\[
	(\beta_s, \theta_s) =\arg\min_{\beta, \theta} L(f_{s-1}(X)+\beta g(X;\theta)),
\]
where
\[
	f_s(X) = f_{s-1}(X) + \beta_s g(X;\theta_s).
\]
The formula above shows that the idea of boosting is similar to forward stagewise regression.
Adaboost is a famous example of such an algorithm with an exponential loss function. The basis function of Adaboost would be the classifier $g(X;\theta) \in \{-1,+1\}$, where $\theta$ represents the parameter of a decision tree.
Let $y_i \in \{-1,+1\}$; then, the loss function of Adaboost can be written as
\begin{align*}
	L(f_s(X)) & = \sum_{n=1}^{N} e^{-y_n f_s(X_n)} \\
			  & = \sum_{n=1}^{N} w_n e^{-y_n\beta g(X_n)}\\
			  & = (e^{\beta}-e^{-\beta}) \sum_{n=1}^{N} w_n \mathbf{I}(y_n \neq g(X_n)) + e^{-\beta}\sum_{n=1}^{N} w_n, 
\end{align*}
where $w_n = e^{-y_n f_{s-1}(X_n)}.$ 
The minimization of the loss function could be divided into two steps. For the first step, the weak learner of the $s$th iteration, $g_s$, is the one that minimizes the weighted errors, i.e.,
\[
	g_s =\arg\min_{g} \sum w_n \mathbf{I} (y_n \neq g(X_n)).
\]

For the second step, we update the $g_s$ in the loss function and minimize the loss to get
\[
	\beta_s = \frac{1}{2}\log\frac{1-\mbox{err}_s}{\mbox{err}_s},
\]

where 
\[
	\mbox{err}_s = \frac{\sum w_n \mathbf{I}(y_n\neq g_s(X_n))}{\sum w_n}.
\]

In the end, the weights $w_{n}(X_n)$, $n=1,2,\cdots, N$, are updated for the next iteration. The whole procedure is presented in Algorithm \ref{adaboost}.
\begin{algorithm}
\caption{Adaboost}\label{adaboost}
\begin{algorithmic}[1]
\State Initial $w = 1/N$ 
\For{$t = 1 \to S$} 
\State{Find a $g_s(X)$ to minimize the weighted errors}
\State{$\mbox{err}_{s} = \frac{\sum w_n \mathbf{I}(y_n \neq g_s)}{\sum w_n}$}
\State{$\beta_s = \frac{1}{2}\log \frac{1-\mbox{err}_s}{\mbox{err}_s}$}
\State $\mbox{Update } w_n \gets w_n\exp(2\beta_s\mathbf{I}(y_n\neq g_s(X_n)))$
\State{Normalize the weights}
\EndFor
\State{$f(X)=sign(\sum_{s=1}^{S} \beta_s g_s(X))$}
\end{algorithmic}
\end{algorithm}


\section{Data} \label{sec:Data}
Progresa is an anti-poverty program in Mexico that provides cash transfer to poor families
to promote investments in children's human capital, such as education, nutrition, and health.
Although the Mexican government initiated compulsory secondary school education (grades 7 to 9) in 1992, the
school attendance rates remain low in rural communities \citep{Bando2004}, where child labor is also a common problem. To improve children's school attendance and evaluate the effect of the conditional cash transfer policy, 
the Progresa program implemented a large-scale, randomized controlled experiment between 1997 and 1999 that covered 24,000 households and 41,000 children whose ages ranged between 6 and 15 \citep{Schultz2004, DeJanvry2006, Todd2006}.

Overall, 506 rural villages participated in the experiment. Of those 506 villages, 186 were randomly assigned to the control group, and 320 of them were assigned to the treatment group. The households in the treatment group were eligible for the cash compensation if their income was below a certain level \citep{Behrman2005}. Of the 24,000 households, 52 percent were eligible to participate in the experiment. An eligible household was to receive a subsidy for each child, from grade 3 through grade 9, that attended school for at least 85 percent of the school days. The amount of subsidy would increase with the child's grade level (Table \ref{table:subsidy}) to offset the opportunity cost of going to school versus working or participating in home production. In addition, the program offered an additional premium for girls because their secondary school attendance rates were lower than those of boys. Overall, the transfer amount was approximately one fourth of the average household income and two thirds that of a child laborer's wage \citep{Parker2000, Schultz2004}. Due to the generosity of the transfer, most of the eligible households participated in the experiment.

\begin{table}[htbp]
  \centering
  \caption{Yearly payment for school attendance in Treatment group}
    \begin{tabularx}{\textwidth}{XXXX}
    \toprule
           &   & \multicolumn{2}{c}{Payment (pesos)}\\
    \cmidrule{3-4}
    Education level & Grade & Boys  & Girls \\
    \midrule
    Primary & 3     & 630    & 630 \\
          & 4     & 720    & 720 \\
          & 5     & 945   & 945 \\
          & 6     & 1215   & 1215 \\
    Secondary & 1     & 1800   & 1890 \\
          & 2     & 1890   & 2115 \\
          & 3     & 2025   & 2295 \\
    \bottomrule
    \end{tabularx}
    \begin{tablenotes}
        \item \emph{Source:} \citet{Schultz2004}, Table 1.
    \end{tablenotes}
  \label{table:subsidy}
\end{table}

To compare the prediction outcome between machine learning algorithms and TW2006, I use the same data set published on the American Economic Review website. The data focus on landless nuclear households, which are usually the poorest ones in the village. This reduces the number of household to 3,201 and the number of children aged 6 to 15 to 4,652 
for the baseline survey conducted in October 1997. The data for the 1997 control group, the 1997 treatment group, and the 1998 control group are aggregated together as training data because the participants in these groups did not receive any cash transfer. The 1998 treatment group alone serve as the test data. There are 14,039 entries\footnote{Each data entry contains information about a child in the eligible household. The age of the child could be younger than 6 or older than 15 years. The information would be used to construct household variables.} in the training data and 5,461 entries in the test data. The entries with missing household income or missing children age data are dropped because they are important for school attendance decisions and because the prediction error of these variables could be very large. This reduces the training data to 12,595 entries and the test data to 4,626 entries and leads to minor difference in school attendance rate between the current study and TW2006 (see Table \ref{tab:attRate}). 

Another difference in the attendance rates comes from the fact that this study does not impose the attendance and work restrictions used by TW2006 to reduce the computational burden.\footnote{\citet{Todd2006} require that a child attend school only if all younger children of the same gender attend school and works for pay only only if all older children of the same gender work for for pay. This restriction drops 6 percent of the data.} Table \ref{tab:attRateSubgroup} shows the attendance rate of each subgroup in both the previous and current studies. Thirty-two subgroups are constructed based on the child's gender, education level, school performance, and experimental design. Since the differences in most subgroups are small and the abandonment is purely due to computational concerns, 
I keep the larger data for better learning.

\begin{table}[htbp]
\begin{threeparttable}
  \centering
  \caption{School Attendance Rate by Age and Gender}
    \begin{tabular}{rrrcrrcrrr}
    \toprule
        &       & \multicolumn{2}{c}{Attendance Rate, TW2006\tnote{a}} &       & \multicolumn{2}{c}{Attendance Rate, Current} 
        &       & \multicolumn{2}{c}{Difference}  \\
		\cmidrule{3-4}\cmidrule{6-7}\cmidrule{9-10}    
		\multicolumn{1}{c}{Age} &       & \multicolumn{1}{c}{Boys} & \multicolumn{1}{c}{Girls} &       & 
		\multicolumn{1}{c}{Boys} & \multicolumn{1}{c}{Girls} &       & 
		\multicolumn{1}{c}{Boys} & \multicolumn{1}{c}{Girls} \\
    \midrule
    6     &       & 92.86  & 95.68  &       & 93.02  & 96.58  &       & 0.16  & 0.90  \\
    7     &       & 98.92  & 97.77  &       & 98.82  & 97.63  &       & -0.10  & -0.14  \\
    8     &       & 98.60  & 99.24  &       & 98.51  & 99.19  &       & -0.09  & -0.05  \\
    9     &       & 99.58  & 99.20  &       & 100.00  & 99.13  &       & 0.42  & -0.07  \\
    10    &       & 97.64  & 98.78  &       & 97.87  & 98.72  &       & 0.23  & -0.06  \\
    11    &       & 98.60  & 96.86  &       & 99.03  & 97.10  &       & 0.43  & 0.24  \\
    12    &       & 88.73  & 90.00  &       & 90.00  & 89.67  &       & 1.27  & -0.33  \\
    13    &       & 78.07  & 70.86  &       & 80.35  & 73.05  &       & 2.28  & 2.19  \\
    14    &       & 67.26  & 60.40  &       & 67.74  & 63.31  &       & 0.48  & 2.91  \\
    15    &       & 47.71  & 40.16  &       & 48.32  & 40.71  &       & 0.61  & 0.55  \\
    \bottomrule
    \end{tabular}%
    \begin{tablenotes}
		\item[a] The attendance rate in \citet{Todd2006}
    \end{tablenotes}
  \label{tab:attRate}%
\end{threeparttable}
\end{table}%

\begin{sidewaystable}[htbp]
\begin{threeparttable}
  \centering
  \caption{School Attendance Rate by Subgroup}
    \begin{tabular}{@{}lll.{1}.{1}c.{1}.{1}c.{1}.{3}c.{1}.{5}@{}}
    \toprule
    \multicolumn{2}{l}{Attendance Rate} &   & 
    \multicolumn{2}{c}{Age 6-11} &   & 
    \multicolumn{2}{c}{Age 12-15} &   & 
    \multicolumn{2}{c}{Age 12-15, behind} &   & 
    \multicolumn{2}{c}{Age 13-15, behind}  \\
      &  &  &  &  &  &  &  &  &  &  &  &\multicolumn{2}{c}{HGC\tnote{a} $\geq$ 6}\\
    \cmidrule{4-5}
    \cmidrule{7-8}
    \cmidrule{10-11}
    \cmidrule{13-14}
      &  &  & 
    \multicolumn{1}{c}{Girls} &\multicolumn{1}{c}{Boys} &  & 
    \multicolumn{1}{c}{Girls} &\multicolumn{1}{c}{Boys} &  & 
    \multicolumn{1}{c}{Girls} &\multicolumn{1}{c}{Boys} &  & 
    \multicolumn{1}{c}{Girls} &\multicolumn{1}{c}{Boys} \\
    \midrule
    \multicolumn{1}{l}{\textit{A. TW2006}} &  &  &  &  &  &  &  &  &  &  &  &\\
    \multicolumn{1}{l}{\mbox{\quad 1997 control}}  
    &  &  &  96.9 &  96.6 &  &  65.3  &  68.8 &  &  58.3 &  64.0 &  & 40.9  & 59.0 \\
	\multicolumn{1}{l}{\mbox{\quad No. of obs.}}   
	&  &  & 449   & 471   &  & 190    & 189   &  & 127   & 139   &  & 66    & 61   \\
    \multicolumn{1}{l}{\mbox{\quad 1997 treatment}}
    &  &  &  97.6 &  97.6 &  &  62.9  &  69.5 &  &  56.9 &  64.2 &  & 30.3  & 52.6 \\
    \multicolumn{1}{l}{\mbox{\quad No. of obs.}}
    &  &  & 632   & 671   &  & 205    & 279   &  & 144   & 204   &  & 66    & 95   \\
    \multicolumn{1}{l}{\mbox{\quad 1998 control}}  
    &  &  &  96.5 &  96.7 &  &  66.5  &  72.5 &  &  58.7 &  67.4 &  & 44.4  & 57.1 \\
    \multicolumn{1}{l}{\mbox{\quad No. of obs.}}   
    &  &  & 431   & 460   &  & 176    & 182   &  & 121   & 135   &  & 72    & 56   \\
    \multicolumn{1}{l}{\mbox{\quad 1998 treatment}}
    &  &  &  98.5 &  98.7 &  &  74.4  &  76.3 &  &  71.4 &  71.6 &  & 51.5  & 58.3 \\
    \multicolumn{1}{l}{\mbox{\quad No. of obs.}}   
    &  &  & 600   & 678   &  & 223    & 262   &  & 161   & 190   &  & 66    & 96   \\
                                 &  &  &  &  &  &  &  &  &  &  &  &\\
    \multicolumn{1}{l}{\textit{B. Current}}  &  &  &  &  &  &  &  &  &  &  &  &\\
    \multicolumn{1}{l}{\mbox{\quad 1997 control}}      
    &  &  &  97.2 &  97.0 &  &  67.7  &  72.8 &  &  57.5 &  66.1 &  & 41.9  &  60.9\\
    \multicolumn{1}{l}{\mbox{\quad No. of obs.}}   
    &  &  & 610   & 602   &  & 257    & 279   &  & 160   & 186   &  & 86    &  87  \\
    \multicolumn{1}{l}{\mbox{\quad 1997 treatment}}  
    &  &  &  98.7 &  98.3 &  &  71.3  &  73.2 &  &  62.6 &  64.0 &  & 41.1  &  50.4\\
    \multicolumn{1}{l}{\mbox{\quad No. of obs.}}   
    &  &  & 825   & 840   &  & 320    & 388   &  & 187   & 250   &  & 90    & 123  \\
    \multicolumn{1}{l}{\mbox{\quad 1998 control}}    
    &  &  &  97.7 &  96.9 &  &  67.3  &  76.9 &  &  57.3 &  70.0 &  & 41.1  &  60.0\\
    \multicolumn{1}{l}{\mbox{\quad No. of obs.}}   
    &  &  & 522   & 522   &  & 217    & 221   &  & 131   & 147   &  & 73    &  65  \\
    \multicolumn{1}{l}{\mbox{\quad 1998 treatment}}  
    &  &  &  98.9 &  99.2 &  &  76.2  &  77.0 &  &  70.7 &  72.2 &  & 53.6  &  58.5\\
    \multicolumn{1}{l}{\mbox{\quad No. of obs.}}   
    &  &  & 729   & 783   &  & 294    & 331   &  & 188   & 223   &  & 84    & 123  \\
    \bottomrule
    \end{tabular}%
    \begin{tablenotes}\small
		\item[a] HGC means highest grade completed.

	\end{tablenotes}
  \label{tab:attRateSubgroup}%
\end{threeparttable}
\end{sidewaystable}%

\section{Model Training}
Following TW2006, two kinds of prediction tasks are used here. The first type is one-step-ahead prediction, which uses training data to construct the models and perform out-of-sample predictions on the test data, i.e., the 1998 treatment group. The second type is $N$-step-ahead prediction. While this is not an out-of-sample prediction, the $N$-step-ahead prediction tries to evaluate the long-term impact by using the school attendance models with initial data to simulate the decision-making processes from the year in which a couple gets married to the last year that the household appears in the data, which is either 1997 or 1998.

The first task requires only a school attendance model, $f(X,C)$, where $X$ is the household characteristics, such as income, distance to school, and parents' attitude toward education, and $C$ is a child's individual characteristics, such as age, grade, and how many years he or she is behind in school. Appendix \ref{app:varInOneStep} lists all the variables in the school attendance model. I use 10-fold cross-validation to select the best parameters for each machine learning model. The training data are divided into 10 groups of equal size. For each group of data, the other 9 groups are used to train the model, and the current group is used as the validation data set. This procedure is repeated 10 times, and each group serves as the validation data set only once. The average error of the validation data set is treated as the criterion for selecting model parameters, such as the depth of the tree and the magnitude of the penalty.

For the $N$-step-ahead prediction task, we require three additional models to perform the simulation:
income, pregnancy, and school failure models. Appendix \ref{app:varInNstep} lists the variables for each model. 
As in the training process for the first task, I use 10-fold cross-validation approach to select the best parameter values for each machine learning model. Then, the model with the fewest within-prediction errors is chosen to perform the simulation task. Table \ref{tab:incMod} compares the within-sample and out-of-sample income predictions of the different models. Random forest has the smallest MAE and RMSE for both the training and test data sets. Therefore, the the long-term simulation uses the random forest model to predict the household income in each period. Table \ref{tab:pregMod} presents the pregnancy predictions for all models. The long-term simulation uses the logit model to determine whether the wife is pregnant in each period because this model has the lowest MAE and RMSE for both within-sample and out-of-sample predictions.
Table \ref{tab:failMod} compares the school failure predictions of the candidate models. As the table shows, there is no model that outperforms the other models in all measurements. Therefore, random forest is chosen to simulate the failure result due to its lowest within-sample errors.

The simulation proceeds as follows. At the beginning of each period, the income model gives the household predicted income in the current period based on the few variables available at the point of decision making, such as age or distance to city. With the predicted income and other household characteristics, the pregnancy model determines whether the wife is pregnant. If the wife is pregnant, the gender is assigned randomly with probabilities equal to the boy-to-girl ratio in the training data. Then, the school attendance model in the one-step-ahead prediction predicts the attendance outcome for each child. For those who go to school, the school failure model then determines whether the children fail their current grade. At the end of the simulation for each period, the information is updated for the next period until each household reaches its final period.

\begin{table}[htbp]
  \centering
  \caption{Model Comparison for Income Prediction}
    \begin{tabularx}{\textwidth}{XXrrXrr}
    \toprule
		Model          &   & \multicolumn{2}{c}{Training} & & \multicolumn{2}{c}{Test}\\
    \cmidrule{3-4}
    \cmidrule{6-7}
                       &   & \multicolumn{1}{c}{MAE} & \multicolumn{1}{c}{RMSE} &
                       & \multicolumn{1}{c}{MAE} & \multicolumn{1}{c}{RMSE}\\
    \midrule
    	TW2006         &   & 7121  & 12663 &  & 6833 & 14257\\
        CART           &   & 6429  & 12526 &  & 6087 & 14112\\
        LASSO          &   & 6424  & 12489 &  & 6061 & 14062\\
        Random forest  &   & 4721  &  9263 &  & 5747 & 13809\\
        Adaboost	   &   & 6188  & 11914 &  & 5891 & 14022\\
    \bottomrule
    \end{tabularx}
    \begin{tablenotes}
		\item[a] The C4.5 algorithm does not work for continuous variables, so the results of C4.5 are omitted.
    \end{tablenotes}
  \label{tab:incMod}
\end{table}

\begin{table}[htbp]
  \centering
  \caption{Model Comparison for Pregnancy Prediction}
    \begin{tabularx}{\textwidth}{XX.{1}.{1}X.{1}.{1}}
    \toprule
		Model          &   & \multicolumn{2}{c}{Training} & & \multicolumn{2}{c}{Test}\\
    \cmidrule{3-4}
    \cmidrule{6-7}
                       &   & \multicolumn{1}{r}{MAE} & \multicolumn{1}{r}{RMSE} &
                       & \multicolumn{1}{r}{MAE} & \multicolumn{1}{r}{RMSE}\\
    \midrule
        CART           &   & 5.02  &  5.55 &  &  3.63 &  6.96\\
        C4.5           &   & 6.75  &  7.39 &  &  1.50 &  1.64\\
        LASSO          &   & 9.42  & 13.42 &  & 11.35 & 16.48\\
        Random forest  &   & 6.08  &  6.57 &  &  4.43 &  6.11\\
        Adaboost	   &   & 6.63  &  7.28 &  &  4.90 &  5.48\\
        Logit          &   & 2.59  &  3.48 &  &  1.30 &  1.45\\
    \bottomrule
    \end{tabularx}
    \begin{tablenotes}
		\item a. Pregnancy is an endogenous choice in TW2006, so the results are not shown in the table.
		\item b. For more prediction details, see Appendix \ref{app:varInNstep}.
    \end{tablenotes}
  \label{tab:pregMod}
\end{table}

\begin{table}[htbp]
  \centering
  \caption{Model Comparison for School Failure Prediction}
    \begin{tabularx}{\textwidth}{XX.{1}.{1}X.{1}.{1}}
    \toprule
		Model          &   & \multicolumn{2}{c}{Training} & & \multicolumn{2}{c}{Test}\\
    \cmidrule{3-4}
    \cmidrule{6-7}
                       &   & \multicolumn{1}{r}{MAE} & \multicolumn{1}{r}{RMSE} &
                       & \multicolumn{1}{r}{MAE} & \multicolumn{1}{r}{RMSE}\\
    \midrule
	    TW2006         &   &  4.36  &   5.87 &  &  1.65 &  2.31\\
        CART           &   &  6.04  &   7.05 &  &  4.20 &  5.14\\
        C4.5           &   & 12.80  &  13.90 &  & 11.64 & 11.88\\
        LASSO          &   & 14.09  &  15.25 &  & 13.31 & 13.48\\
        Random forest  &   &  3.33  &   4.24 &  &  9.41 & 10.67\\
        Adaboost	   &   &  7.94  &  10.30 &  &  6.98 &  8.32\\
        Logit          &   &  4.78  &   6.40 &  &  3.44 &  5.81\\
    \bottomrule
    \end{tabularx}
  \label{tab:failMod}
\end{table}

\section{Results}

\subsection{One-Step-Ahead Prediction}
Tables \ref{tab:OnePeriodPrediction} and \ref{tab:OneStepErr} show the out-of-sample forecast for TW2006 and the machine learning models. As panel A of Table \ref{tab:OnePeriodPrediction} shows, the original structural model performs extremely well for the 1998 treatment group, except for the last two subgroups (the last two columns in Table \ref{tab:OnePeriodPrediction}). For the first two subgroups, boys and girls who are between 6 to 11 years old, more than 98 percent of children attend school, and the structural model also predicts high attendance rates for them. The school attendance rates drop significantly when children are between 12 and 15 years old, going from 98 percent for the first two subgroups to 70 percent for the third to sixth subgroups. The predictions of the structural model are still very close to the actual outcomes for these subgroups, with prediction errors of approximately one percentage point. The attendance rates drop further when the teenagers complete their primary school education. In the last two columns of Table \ref{tab:OnePeriodPrediction}, the attendance rate is 51.5 percent for girls and 58.3 percent for boys in that category. In this category, the structural model overpredicts girls' attendance rates by 7.2 percentage points and boys' rates by 8.4 percentage points. The MAE of the attendance rate for the structural model is 2.76 percentage points.

Panel B of Table \ref{tab:OnePeriodPrediction} shows the predictions of the machine learning models for the treatment group. All machine learning models have smaller MAEs than the structural model, as shown in Table \ref{tab:OneStepErr}. The result remains the same even after we exclude the first two subgroups and focus only on children aged 12 to 15. The ranking of all models in terms of the MAE, from best to worst, is Adaboost, random forest, LASSO, CART, C4.5, and the structural model. According to the RMSEs, this ranking is Adaboost, LASSO, random forest, CART, C4.5, and the structural model. Overall, the machine learning models tend to perform better in the first and last two subgroups but perform worse in the middle subgroups. The improvements for the first two subgroups are small because the structural model yields a close prediction, but the improvements for the last two subgroups are significant. Most of the models reduce the prediction errors by more than 4 percentage points in the last category. The models, except C4.5, seem to overpredict the attendance rates in the first five subgroups and underpredict the attendance rates in the last three subgroups.

If we look at each individual model, CART tends to underpredict teenage boys' attendance rate and overpredict teenage girls' attendance rate. The largest prediction error of the CART model, 5.7 percentage points, also occurs in the last subgroup. Figure \ref{fig:CART_train} in Appendix \ref{app:ML_model} presents the decision tree of CART constructed from the training data. The decision tree captures the facts that children younger than 6 years old never go to school and that children between 6 and 12 years old almost always attend school. It uses only six variables to build the tree, and the children's age and the number of years by which they are behind in school play an important role in determining whether a child attends school. However, the CART model does not capture the influence of household income, which means that the model does not reflect any changes induced by the CCT.

\begin{sidewaystable}[htbp]
  \centering
  \caption{One-Step-Ahead Prediction for 1998 Treatment Group}
    \begin{tabular}{@{}lll.{1}.{1}c.{1}.{1}c.{1}.{3}c.{1}.{5}@{}}
    \toprule
        \multicolumn{2}{l}{Attendance Rate} &   & \multicolumn{2}{c}{Age 6-11} &   & \multicolumn{2}{c}{Age 12-15} &   & 			        \multicolumn{2}{c}{Age 12-15, behind} &   & \multicolumn{2}{c}{Age 13-15, behind}  \\
        &  &  &  &  &  &  &  &  &  &  &  &  \multicolumn{2}{c}{HGC* $\geq$ 6}\\
    \cmidrule{4-5}
    \cmidrule{7-8}
    \cmidrule{10-11}
    \cmidrule{13-14}
        &  &  & \multicolumn{1}{c}{Girls} & \multicolumn{1}{c}{Boys} & & 
        \multicolumn{1}{c}{Girls} & \multicolumn{1}{c}{Boys} & & 
        \multicolumn{1}{c}{Girls} & \multicolumn{1}{c}{Boys} & & 
        \multicolumn{1}{c}{Girls} & \multicolumn{1}{c}{Boys} \\
    \midrule
    \multicolumn{1}{l}{\textit{A. TW2006}} &  &  &  &  &  &  &  &  &  &  &  &\\
                             & Actual       &   & 98.5  & 98.7  &   & 74.4  & 76.3  &   & 71.4  & 71.6  &   & 51.5  & 58.3 \\
                             & Predicted    &   & 97.1  & 97.1  &   & 74.9  & 77.1  &   & 72.3  & 72.9  &   & 58.7  & 66.7 \\
                             & Err          &   & -1.4  & -1.6  &   & 0.5   & 0.8   &   & 0.9   & 1.3   &   & 7.2   & 8.4 \\
                                            &  &  &  &  &  &  &  &  &  &  &  &\\
	\multicolumn{1}{l}{\textit{B. Current}} &  &  &  &  &  &  &  &  &  &  &  &\\                             
                             & Actual       &   & 98.9  & 99.2  &   & 76.2  & 77.0  &   & 70.7  & 72.2  &   & 53.6  & 58.5 \\
    \multicolumn{1}{l}{\mbox{\quad CART}}      
    						 &              &   &       &       &   &       &       &   &       &       &   &       &      \\                             
                             & Predicted    &   & 100   & 100   &   & 80.3  & 77.6  &   & 73.4  & 68.2  &   & 52.4  & 52.8\\
                             & Err          &   & 1.1   & 0.8   &   &  4.1  &  0.6  &   &  2.7  & -4.0  &   & -1.2  & -5.7 \\
	\multicolumn{1}{l}{\mbox{\quad C4.5}}       
    						 &              &   &       &       &   &       &       &   &       &       &   &       &     \\
                             & Predicted    &   & 98.6  & 97.7  &   & 73.5  & 71.9  &   & 69.1  & 65.5  &   & 51.2  & 58.5 \\
                             & Err          &   & -0.3  & -1.5  &   & -2.7  & -5.1  &   & -1.6  & -6.7  &   & -2.4  &  0.0 \\
    \multicolumn{1}{l}{\mbox{\quad LASSO}}      
    						 &              &   &       &       &   &       &       &   &       &       &   &       &     \\
                             & Predicted    &   & 99.7  & 99.7  &   & 79.6  & 75.8  &   & 72.9  & 68.2  &   & 56.0  & 54.5 \\
                             & Err          &   &  0.8  &  0.5  &   &  3.4  & -1.2  &   &  2.2  & -4.0  &   &  2.4  & -4.0 \\
	\multicolumn{1}{l}{\mbox{\quad Random Forest}}      
    						 &              &   &       &       &   &       &       &   &       &       &   &       &     \\
                             & Predicted    &   & 99.9  & 100   &   & 82.0  & 80.4  &   & 75.0  & 71.3  &   & 52.4  & 58.5 \\
                             & Err          &   &  1.0  & 0.8   &   &  5.8  &  3.4  &   &  4.3  & -0.9  &   & -1.2  &  0.0 \\
	\multicolumn{1}{l}{\mbox{\quad Adaboost}}      
    						 &           &   &       &       &   &       &       &   &       &       &   &       &     \\
                             & Predicted    &   & 100   & 100   &   & 81.6  & 79.2  &   & 72.9  & 69.1  &   & 53.6  & 56.1 \\
                             & Err          &   & 1.1   & 0.8   &   &  5.4  &  2.2  &   &  2.2  & -3.1  &   &  0.0  & -2.4 \\
    \bottomrule
    \end{tabular}%
   \label{tab:OnePeriodPrediction}
      \begin{tablenotes}\small
        \item * HGC, the highest completed grade.
    \end{tablenotes}
\end{sidewaystable}%

C4.5 has the worst prediction performance of all the machine learning algorithms. The model tends to underpredict the attendance rate for almost every subgroup. The largest error occurs in the sixth subgroup and is 6.7 percentage points for teenage girls who are behind in school. It is also the largest prediction error of all the machine learning predictions. The MAE and RMSE of the C4.5 model are close to those of the CART model, which is not a surprise because C4.5 is an improved CART algorithm. C4.5 also does not capture the effect of CCT in its decision tree model\footnote{The decision tree of C4.5 is much more complicated than that of CART model. Therefore I do not present it here.}. The prediction errors of the LASSO model are all within four percentage points, and the mean absolute error is 2.3 percentage points. LASSO also tends to overpredicts teenage girls' attendance rate and underpredicts teenage boys' attendance rate, similar to CART. Appendix \ref{app:ML_model} shows the estimated parameters of the LASSO model.

The forecasts for the random forest model are also close to the actual outcomes. The mean absolute error is 2.18 percentage points, which is very close to that of the best model, Adaboost. The largest prediction errors for the random forest model occur in the third through fifth subgroups. For other subgroups, the prediction errors are approximately 1 percentage point. Figure \ref{fig:rfVar} in Appendix \ref{app:ML_model} presents the importance of each variable, as measured by its impact on the accuracy or Gini gain of the out-of-bag samples. Adaboost has the best performance in terms of MAE and RMSE. Its average absolute prediction error is 2.15 percentage points, and the largest prediction occurs in the third subgroup, which is 5.4 percentage points.


\begin{table}[htbp]
  \centering
  \caption{Model Comparison for One-Step-Ahead Prediction: MAE and RMSE}
    \begin{tabularx}{\textwidth}{XXXX}
    \toprule
     	Model          &   & MAE   & RMSE\\
    \midrule
    	TW2006         &   & 2.76  & 4.04 \\
        CART           &   & 2.53  & 3.09 \\
        C4.5           &   & 2.54  & 3.32\\
        LASSO          &   & 2.31  & 2.66 \\
        Random forest  &   & 2.18  & 2.91 \\
        Adaboost	   &   & 2.15  & 2.65 \\
    \bottomrule
    \end{tabularx}
  \label{tab:OneStepErr}
\end{table}

\subsection{Model Accuracy}

In addition to the MAE and RMSE comparisons, I calculate the accuracy of each machine learning model to measure their performance from another perspective. The accuracy measurement is very common in the machine learning literature for binary choice problems since the false positive and false negative predictions can cancel each other out, thus distorting the evaluation of the model. Table \ref{tab:Acc} compares the accuracy of the five machine learning models for each subgroup.
The accuracy tends to decrease as the restriction increases, which could be caused by the decreasing sample sizes. As the table shows, the prediction error does not directly reveal the underlying accuracy. For example, although C4.5's prediction error for the sixth subgroup, 6.7 percentage points, is larger than the zero percentage point error in the last subgroup, the accuracy this subgroup achieves is 76.23 percent, which is higher than the 72.36 percent achieved for the last subgroup. The same phenomenon happens for between-model comparisons.
The predictions of the C4.5 and random forest algorithms for the last subgroup are exactly the same as the actual outcome; the accuracies, however, are 72.36 percent and 80.48 percent, respectively.

If we rank the models by their accuracy, random forest and Adaboost remain the best two, followed by CART, while LASSO drops to fourth and C4.5 remains the worst model. For all machine learning models, the accuracy for the first two groups is very close to 100 percent because almost every child between 6 and 11 years old goes to school, which simplifies the prediction problem. For the third and fourth subgroups, the accuracies of random forest, Adaboost, and CART are approximately 85 percent, that of Lasso is approximately 80 percent, and that of C4.5 is between 75 and 80 percent. The accuracies remain at the same level for most models in the fifth and sixth subgroups. The only exception is CART, whose accuracy drops to 80 percent, which is the level of LASSO's accuracy. The last two subgroups are the most challenging for prediction as the accuracies of all models decreases significantly, especially for the girls' attendance rate prediction. The accuracies of most models drop at least six percent for the girls' prediction. Random forest and Adaboost achieve 76 percent accuracy for girls and approximately 80 percent for boys. The accuracy of CART is close to those of random forest and Adaboost for boys but is much worse for girls. LASSO and C4.5 achieve 65 percent accuracy for girls and below 80 percent for boys.

\begin{sidewaystable}[htbp]
  \centering
  \caption{Accuracy of One-Step-Ahead Prediction}
    \begin{tabular}{llrrrrrrrrrrrrrr}
    \toprule
        \multicolumn{2}{l}{Attendance Rate}   &   & \multicolumn{2}{c}{Age 6-11}          &   & 
        \multicolumn{2}{c}{Age 12-15}         &   & \multicolumn{2}{c}{Age 12-15, behind} &   & 
        \multicolumn{2}{c}{Age 13-15, behind} &   & \multicolumn{1}{c}{Average }\\
                       &                      &   &                    &                  &   &                                       					   &                      &   &                    &                  &   &   
        \multicolumn{2}{c}{HGC* $\geq$ 6}     &   &\\
    \cmidrule{4-5}
    \cmidrule{7-8}
    \cmidrule{10-11}
    \cmidrule{13-14}
                       &                                 &  & \multicolumn{1}{c}{Girls} & \multicolumn{1}{c}{Boys} &  & 
    \multicolumn{1}{c}{Girls} & \multicolumn{1}{c}{Boys} &  & \multicolumn{1}{c}{Girls} & \multicolumn{1}{c}{Boys} &  &       
    \multicolumn{1}{c}{Girls} & \multicolumn{1}{c}{Boys} &  &\\
    \midrule
    \multicolumn{1}{l}{CART} &         &  & 98.90 & 99.23 &  & 85.03 & 84.29 &  & 81.38 & 83.41 &  & 70.24 & 79.67 &  & 85.27\\
    \multicolumn{1}{l}{C4.5} &         &  & 97.81 & 96.93 &  & 75.51 & 76.74 &  & 75.00 & 76.23 &  & 66.67 & 72.36 &  & 79.66\\
    \multicolumn{1}{l}{LASSO}&         &  & 98.63 & 98.98 &  & 77.55 & 79.46 &  & 76.60 & 78.92 &  & 64.29 & 76.42 &  & 81.36\\
    \multicolumn{1}{l}{Random Forest}& &  & 98.77 & 99.23 &  & 85.37 & 86.40 &  & 84.04 & 85.65 &  & 77.38 & 80.48 &  & 87.17\\
    \multicolumn{1}{l}{Adaboost}&      &  & 98.90 & 99.23 &  & 85.03 & 85.50 &  & 84.04 & 85.20 &  & 76.19 & 82.93 &  & 87.13\\
    \bottomrule
    \end{tabular}%
   \label{tab:Acc}
\end{sidewaystable}%

Tables \ref{tab:OneStep97control}, \ref{tab:OneStep98control}, \ref{tab:OneStep97treat}, and \ref{tab:OneStepErr2} present the within-sample fitness of the structural and machine learning models. Although all machine learning models perform better than does the structural model in terms of out-of-sample forecasts, only the C4.5 and random forest models have better within-sample fitness. This is quite common in the prediction literature because a model with extremely good within-sample fitness tends to be too specific to the current data set and hence lose its generalization ability. 

\begin{sidewaystable}[htbp]
  \centering
  \caption{Within Sample Fitness for 1997 Control Group}
    \begin{tabular}{@{}lll.{1}.{1}c.{1}.{1}c.{1}.{3}c.{1}.{5}@{}}
    \toprule
        \multicolumn{2}{l}{Attendance Rate} &   & \multicolumn{2}{c}{Age 6-11} &   & \multicolumn{2}{c}{Age 12-15} &   & 					\multicolumn{2}{c}{Age 12-15, behind} &   & \multicolumn{2}{c}{Age 13-15, behind}  \\
        &  &  &  &  &  &  &  &  &  &  &  &  \multicolumn{2}{c}{HGC* $\geq$ 6}\\
    \cmidrule{4-5}
    \cmidrule{7-8}
    \cmidrule{10-11}
    \cmidrule{13-14}
        &  &  & \multicolumn{1}{c}{Girls} & \multicolumn{1}{c}{Boys} & & \multicolumn{1}{c}{Girls} & 
        \multicolumn{1}{c}{Boys} & & \multicolumn{1}{c}{Girls} & \multicolumn{1}{c}{Boys} & & 
        \multicolumn{1}{c}{Girls} & \multicolumn{1}{c}{Boys} \\
    \midrule
    \multicolumn{1}{l}{\textit{A. TW2006}} &  &  &  &  &  &  &  &  &  &  &  &\\
                             & Actual       &   & 96.9  & 96.6  &   & 65.3  & 68.8  &   & 58.3  & 64.0  &   & 40.9  & 59.0 \\
                             & Predicted    &   & 96.1  & 96.4  &   & 61.6  & 68.8  &   & 54.2  & 63.9  &   & 40.2  & 55.0 \\
                             & Err          &   & -0.8  & -0.2  &   & -3.7  &  0.0  &   & -4.1  & -0.1  &   & -0.7  & -4.0 \\
                                            &  &  &  &  &  &  &  &  &  &  &  &\\
	\multicolumn{1}{l}{\textit{B. Current}} &  &  &  &  &  &  &  &  &  &  &  &\\                             
							 & Actual       &   & 97.2  & 97.0  &   & 67.7  & 72.8  &   & 57.5  & 66.1  &   & 41.9  & 60.9 \\
    \multicolumn{1}{l}{CART} &              &   &       &       &   &       &       &   &       &       &   &       &      \\
                             & Predicted    &   & 100   & 100   &   & 71.6  & 78.5  &   & 58.8  & 70.4  &   & 39.5  & 57.5 \\
                             & Err          &   &  2.8  &  3.0  &   &  3.9  &  5.7  &   &  1.3  &  4.3  &   & -2.4  & -3.4 \\
    \multicolumn{1}{l}{C4.5} &              &   &       &       &   &       &       &   &       &       &   &       &     \\
                             & Predicted    &   & 97.4  & 97.8  &   & 68.1  & 73.8  &   & 58.1  & 67.7  &   & 43.0  & 63.2 \\
                             & Err          &   &  0.2  &  0.8  &   &  0.4  &  1.0  &   &  0.6  &  1.6  &   &  1.1  &  2.3\\
    \multicolumn{1}{l}{LASSO}&              &   &       &       &   &       &       &   &       &       &   &       &     \\
                             & Predicted    &   & 98.9  & 99.3  &   & 73.5  & 73.5  &   & 61.3  & 65.1  &   & 46.5  & 50.6 \\
                             & Err          &   &  1.7  &  2.3  &   &  5.8  &  0.7  &   &  3.8  & -1.0  &   &  4.6  &-10.3\\
    \multicolumn{1}{l}{Random Forest}       &  &  &  &  &  &  &  &  &  &  &  &  &     \\
                             & Predicted    &   & 98.4  & 98.2  &   & 68.1  & 74.6  &   & 57.5  & 67.2  &   & 41.9  & 60.9 \\
                             & Err          &   &  1.2  &  1.2  &   &  0.4  &  1.8  &   &  0.0  &  1.1  &   &  0.0  &  0.0 \\
    \multicolumn{1}{l}{AdaBoost}&           &   &       &       &   &       &       &   &       &       &   &       &     \\
                             & Predicted    &   & 99.3  & 99.3  &   & 71.6  & 77.8  &   & 57.5  & 67.7  &   & 38.4  & 56.3 \\
                             & Err          &   &  2.1  &  2.3  &   &  3.9  &  5.0  &   &  0.0  &  1.6  &   & -3.5  & -4.6 \\
    \bottomrule
    \end{tabular}%
   \label{tab:OneStep97control}
\end{sidewaystable}%

\begin{sidewaystable}[htbp]
  \centering
  \caption{Within Sample Fitness for 1998 Control Group}
    \begin{tabular}{@{}lll.{1}.{1}c.{1}.{1}c.{1}.{3}c.{1}.{5}@{}}
    \toprule
        \multicolumn{2}{l}{Attendance Rate} &   & \multicolumn{2}{c}{Age 6-11} &   & \multicolumn{2}{c}{Age 12-15} &   & 					\multicolumn{2}{c}{Age 12-15, behind} &   & \multicolumn{2}{c}{Age 13-15, behind}  \\
        &  &  &  &  &  &  &  &  &  &  &  &  \multicolumn{2}{c}{HGC* $\geq$ 6}\\
    \cmidrule{4-5}
    \cmidrule{7-8}
    \cmidrule{10-11}
    \cmidrule{13-14}
        &  &  & \multicolumn{1}{c}{Girls} & \multicolumn{1}{c}{Boys} & & \multicolumn{1}{c}{Girls} & 
        \multicolumn{1}{c}{Boys} & & \multicolumn{1}{c}{Girls} & \multicolumn{1}{c}{Boys} & & 
        \multicolumn{1}{c}{Girls} & \multicolumn{1}{c}{Boys} \\
    \midrule
    \multicolumn{1}{l}{\textit{A. TW2006}} &  &  &  &  &  &  &  &  &  &  &  &\\
                             & Actual       &   & 96.5  & 96.7  &   & 66.5  & 72.5  &   & 58.7  & 67.4  &   & 44.4  & 57.1 \\
                             & Predicted    &   & 96.2  & 96.4  &   & 61.8  & 68.8  &   & 55.5  & 65.3  &   & 45.3  & 53.0  \\
                             & Err          &   & -0.3  & -0.3  &   & -4.7  & -3.7  &   & -3.2  & -2.1  &   &  0.9  & -4.1  \\
                                            &  &  &  &  &  &  &  &  &  &  &  &\\
	\multicolumn{1}{l}{\textit{B. Current}} &  &  &  &  &  &  &  &  &  &  &  &\\                             
							 & Actual       &   & 97.7  & 96.9  &   & 67.3  & 76.9  &   & 57.3  & 70.0  &   & 41.1  & 60.0 \\
    \multicolumn{1}{l}{CART} &              &   &       &       &   &       &       &   &       &       &   &       &      \\
                             & Predicted    &   & 100   & 100   &   & 73.7  & 86.4  &   & 63.4  & 81.0  &   & 45.2  & 67.7 \\
                             & Err          &   & 2.3   & 3.1   &   &  6.4  &  9.5  &   &  6.1  & 11.0  &   & 4.1   &  7.7 \\
    \multicolumn{1}{l}{C4.5} &              &   &       &       &   &       &       &   &       &       &   &       &     \\
                             & Predicted    &   & 97.9  & 97.5  &   & 66.8  & 78.3  &   & 56.5  & 72.1  &   & 41.1  & 60.0 \\
                             & Err          &   &  0.2  &  0.6  &   & -0.5  &  1.4  &   & -0.8  &  2.1  &   &  0.0  &  0.0\\
    \multicolumn{1}{l}{LASSO}&              &   &       &       &   &       &       &   &       &       &   &       &     \\
                             & Predicted    &   & 99.0  & 98.7  &   & 78.3  & 82.4  &   & 67.2  & 75.5  &   & 52.1  & 64.6 \\
                             & Err          &   &  1.3  &  1.8  &   & 11.0  &  5.5  &   &  9.9  &  5.5  &   & 11.0  &  4.6\\
    \multicolumn{1}{l}{Random Forest}       &  &  &  &  &  &  &  &  &  &  &  &  &     \\
                             & Predicted    &   & 98.9  & 99.2  &   & 68.7  & 79.2  &   & 58.0  & 72.1  &   & 42.5  & 61.5 \\
                             & Err          &   &  1.2  &  2.3  &   &  1.4  &  2.3  &   &  0.7  &  2.1  &   &  1.4  &  1.5 \\
    \multicolumn{1}{l}{AdaBoost}&           &   &       &       &   &       &       &   &       &       &   &       &     \\
                             & Predicted    &   & 99.0  & 99.8  &   & 73.7  & 82.4  &   & 60.3  & 74.8  &   & 46.6  & 58.5 \\
                             & Err          &   &  1.3  &  2.9  &   &  6.4  &  5.5  &   &  3.0  &  4.8  &   &  5.5  & -1.5 \\
    \bottomrule
    \end{tabular}%
   \label{tab:OneStep98control}
\end{sidewaystable}%

\begin{sidewaystable}[htbp]
  \centering
  \caption{Within Sample Fitness for 1997 Treatment Group}
    \begin{tabular}{@{}lll.{1}.{1}c.{1}.{1}c.{1}.{3}c.{1}.{5}@{}}
    \toprule
        \multicolumn{2}{l}{Attendance Rate} &   & \multicolumn{2}{c}{Age 6-11} &   & \multicolumn{2}{c}{Age 12-15} &   & 					\multicolumn{2}{c}{Age 12-15, behind} &   & \multicolumn{2}{c}{Age 13-15, behind}  \\
        &  &  &  &  &  &  &  &  &  &  &  &  \multicolumn{2}{c}{HGC* $\geq$ 6}\\
    \cmidrule{4-5}
    \cmidrule{7-8}
    \cmidrule{10-11}
    \cmidrule{13-14}
        &  &  & \multicolumn{1}{c}{Girls} & \multicolumn{1}{c}{Boys} & & \multicolumn{1}{c}{Girls} & 
        \multicolumn{1}{c}{Boys} & & \multicolumn{1}{c}{Girls} & \multicolumn{1}{c}{Boys} & & 
        \multicolumn{1}{c}{Girls} & \multicolumn{1}{c}{Boys} \\
    \midrule
    \multicolumn{1}{l}{\textit{A. TW2006}} &  &  &  &  &  &  &  &  &  &  &  &\\
                             & Actual       &   & 97.6  & 97.6  &   & 62.9  & 69.5  &   & 56.9  & 64.2  &   & 30.3  & 52.6 \\
                             & Predicted    &   & 96.4  & 96.3  &   & 61.8  & 68.0  &   & 55.6  & 62.7  &   & 37.3  & 51.7  \\
                             & Err          &   & -1.2  & -1.3  &   & -1.1  & -1.5  &   & -1.3  & -1.5  &   &  7.0  & -0.9  \\
                                            &  &  &  &  &  &  &  &  &  &  &  &\\
	\multicolumn{1}{l}{\textit{B. Current}} &  &  &  &  &  &  &  &  &  &  &  &\\                             
							 & Actual       &   & 98.7  & 98.3  &   & 71.3  & 73.2  &   & 62.6  & 64.0  &   & 41.1  & 50.4 \\
    \multicolumn{1}{l}{CART} &              &   &       &       &   &       &       &   &       &       &   &       &      \\
                             & Predicted    &   & 100   & 100   &   & 78.4  & 80.2  &   & 65.8  & 70.0  &   & 44.4  & 53.7 \\
                             & Err          &   & 1.3   & 1.7   &   &  7.1  &  7.0  &   &  3.2  &  6.0  &   &  3.3  &  3.3\\
    \multicolumn{1}{l}{C4.5} &              &   &       &       &   &       &       &   &       &       &   &       &     \\
                             & Predicted    &   & 98.9  & 98.9  &   & 71.3  & 72.7  &   & 63.6  & 62.8  &   & 40.0  & 48.0 \\
                             & Err          &   &  0.2  &  0.6  &   &  0.0  & -0.5  &   &  1.0  & -1.2  &   & -1.1  & -2.4 \\
    \multicolumn{1}{l}{LASSO}&              &   &       &       &   &       &       &   &       &       &   &       &     \\
                             & Predicted    &   & 98.5  & 99.4  &   & 75.9  & 75.5  &   & 63.1  & 65.2  &   & 43.3  & 51.2 \\
                             & Err          &   & -0.2  &  1.1  &   &  4.6  &  2.3  &   &  0.5  &  1.2  &   &  2.2  &  0.8 \\
    \multicolumn{1}{l}{Random Forest}       &  &  &  &  &  &  &  &  &  &  &  &  &     \\
                             & Predicted    &   & 99.5  & 99.5  &   & 73.1  & 74.2  &   & 64.2  & 64.4  &   & 41.1  & 50.4 \\
                             & Err          &   &  0.8  &  1.2  &   &  1.8  &  1.0  &   &  1.6  &  0.4  &   &  0.0  &  0.0 \\
    \multicolumn{1}{l}{AdaBoost}&           &   &       &       &   &       &       &   &       &       &   &       &     \\
                             & Predicted    &   & 99.6  & 99.9  &   & 79.1  & 79.4  &   & 65.8  & 68.8  &   & 47.8  & 54.5 \\
                             & Err          &   &  0.9  &  1.6  &   &  7.8  &  6.2  &   &  3.2  &  4.8  &   &  6.7  &  4.1\\
    \bottomrule
    \end{tabular}%
   \label{tab:OneStep97treat}
\end{sidewaystable}%

\begin{table}[htbp]
  \centering
  \caption{Model Comparison for One-Step-Ahead Prediction: MAE and RMSE}
    \begin{tabularx}{\textwidth}{XXXXX}
    \toprule
     	Model          &   & MAE   & RMSE & Accuracy\\
    \midrule
    	TW2006         &   & 2.03  & 2.71 & NA\\
        CART           &   & 4.58  & 5.22 & 86.23\\
        C4.5           &   & 0.86  & 1.10 & 97.87\\
        LASSO          &   & 3.90  & 5.19 & 81.09\\
        Random forest  &   & 1.06  & 1.28 & 98.95\\
        Adaboost	   &   & 3.72  & 4.23 & 89.30\\
    \bottomrule
    \end{tabularx}
  \label{tab:OneStepErr2}
\end{table}

\subsection{N-step ahead prediction}
The $N$-step ahead prediction uses only variables that are available at the time the household formed to simulate the long-term decision making process from the year a couple gets married to the last year they appear in the data, either 1997 or 1998.
Tables \ref{tab:NStep97control}, \ref{tab:NStep98control}, and \ref{tab:NStep97treat} show the comparison of the machine learning and structural models.

The prediction errors of all models are larger than the errors in the out-of-sample forecast because there is less information available in each period and the fact that the prediction error accumulates over time. However, the structural model manages to restrict the prediction error to approximately 10 percentage points, while all the machine learning models have very large errors. The cause of the large machine learning model errors in the long-term simulation could be the result of inaccuracies in the income, pregnancy, and school failure prediction models. 

Table \ref{tab:incErr} in Appendix \ref{app:varInNstep} shows the income predictions of TW2006 and the current model (random forest in this case) across the different subgroups for the current data set. Although the random forest-based model has a smaller MAE and RMSE than TW2006 does in all subgroups, the values are still very large compared to the amount of the subsidy. The magnitude of the MAE and RMSE can be several times the size of the cash transfer. Since previous studies have shown that subsidies, which affect household income, have a significant impact on school attendance, the large error in income prediction must greatly affect the simulation quality. 

Table \ref{tab:pregErr} in Appendix \ref{app:varInNstep} presents the pregnancy prediction of the logit model in different age groups for the current data. As Table \ref{tab:logitPregCoef} shows, the effect of household income is significant; hence, the pregnancy prediction is also affected by any inaccuracy in the income prediction. The school failure prediction model might be another reason for the poor fitness in the long-term simulation. As Figure \ref{fig:CART_train}, Table \ref{tab:lassoCoef}, and Figure \ref{fig:rfVar} show, the number of years that a child is behind in school is a key factor for school attendance. Table \ref{tab:failErr} compares the school failure predictions of TW2006 and the random forest model. Again, although the within-sample MAE and RMSE of the random forest model are smaller than those of the structural model, as shown in Table \ref{tab:failMod}, the prediction errors are still quite large in many subgroups compared to the actual failure rate.

The long-term within-sample simulation reveals the different properties between the machine learning and structural models. When either the sample size or the number of covariates is limited, the predictive power of the machine learning models is also limited. However, the structural econometric model might be able to overcome this problem by imposing constraints on the choice set or the distribution of unobservable variables. Therefore, the structural model could complement machine learning models when data are difficult or expensive to acquire.

\begin{sidewaystable}[htbp]
  \centering
  \caption{$N$-Step-Ahead Simulation, 1997 Control Group}
    \begin{tabular}{@{}lll.{1}.{1}c.{1}.{1}c.{1}.{3}c.{1}.{5}@{}}
    \toprule
        \multicolumn{2}{l}{Attendance Rate} &   & \multicolumn{2}{c}{Age 6-11} &   & \multicolumn{2}{c}{Age 12-15} &   & 					\multicolumn{2}{c}{Age 12-15, behind} &   & \multicolumn{2}{c}{Age 13-15, behind}  \\
        &  &  &  &  &  &  &  &  &  &  &  &  \multicolumn{2}{c}{HGC* $\geq$ 6}\\
    \cmidrule{4-5}
    \cmidrule{7-8}
    \cmidrule{10-11}
    \cmidrule{13-14}
        &  &  & \multicolumn{1}{c}{Girls} & \multicolumn{1}{c}{Boys} & & \multicolumn{1}{c}{Girls} & 
        \multicolumn{1}{c}{Boys} & & \multicolumn{1}{c}{Girls} & \multicolumn{1}{c}{Boys} & & 
        \multicolumn{1}{c}{Girls} & \multicolumn{1}{c}{Boys} \\
    \midrule
    \multicolumn{1}{l}{\textit{A. TW2006}} &  &  &  &  &  &  &  &  &  &  &  &\\
                             & Actual       &   & 96.9  & 96.6  &   & 65.3  & 68.8  &   & 58.3  & 64.0  &   & 40.9  & 59.0 \\
                             & Predicted    &   & 95.3  & 93.3  &   & 58.2  & 62.5  &   & 52.4  & 56.4  &   & 41.3  & 51.1 \\
                             & Err          &   & -1.6  & -3.3  &   & -7.1  & -6.3  &   & -5.9  & -7.6  &   &  0.4  & -7.9 \\
                                            &  &  &  &  &  &  &  &  &  &  &  &\\
	\multicolumn{1}{l}{\textit{B. Current}} &  &  &  &  &  &  &  &  &  &  &  &\\                             
							 & Actual       &   & 97.2  & 97.0  &   & 67.7  & 72.8  &   & 57.5  & 66.1  &   & 41.9  & 60.9 \\
    \multicolumn{1}{l}{CART} &              &   &       &       &   &       &       &   &       &       &   &       &      \\
                             & Predicted    &   & 100   & 100   &   & 84.9  & 77.0  &   & 66.7  & 69.9  &   & 65.4  & 22.2 \\
                             & Err          &   &  2.8  &  3.0  &   & 17.2  &  4.2  &   &  9.2  &  3.8  &   & 23.5  &-38.7 \\
    \multicolumn{1}{l}{C4.5} &              &   &       &       &   &       &       &   &       &       &   &       &     \\
                             & Predicted    &   & 97.0  & 94.6  &   & 72.5  & 75.9  &   & 50.6  & 61.0  &   & 48.6  & 48.1 \\
                             & Err          &   & -0.2  & -2.4  &   &  4.8  &  3.1  &   & -6.9  & -5.1  &   &  6.7  &-12.8 \\
    \multicolumn{1}{l}{LASSO}&              &   &       &       &   &       &       &   &       &       &   &       &     \\
                             & Predicted    &   & 95.8  & 94.6  &   & 80.2  & 80.7  &   & 60.6  & 64.0  &   & 63.0  & 52.5 \\
                             & Err          &   & -1.4  & -2.4  &   & 12.5  &  7.9  &   &  3.1  & -2.1  &   & 21.1  & -8.4\\
    \multicolumn{1}{l}{Random Forest}       &  &  &  &  &  &  &  &  &  &  &  &  &     \\
                             & Predicted    &   & 100   & 99.6  &   & 83.1  & 84.5  &   & 62.5  & 67.4  &   & 50.0  & 28.0 \\
                             & Err          &   &  2.8  &  2.6  &   & 15.4  & 11.7  &   &  5.0  &  1.3  &   &  8.1  &-32.9 \\
    \multicolumn{1}{l}{AdaBoost}&           &   &       &       &   &       &       &   &       &       &   &       &     \\
                             & Predicted    &   & 100   & 99.8  &   & 84.0  & 85.9  &   & 61.5  & 71.0  &   & 51.9  & 40.7 \\
                             & Err          &   &  2.8  &  2.8  &   & 16.3  & 13.1  &   &  4.0  &  4.9  &   & 10.0  &-20.2\\
    \bottomrule
    \end{tabular}%
   \label{tab:NStep97control}
\end{sidewaystable}%

\begin{sidewaystable}[htbp]
  \centering
  \caption{$N$-Step-Ahead Simulation, 1998 Control Group}
    \begin{tabular}{@{}lll.{1}.{1}c.{1}.{1}c.{1}.{3}c.{1}.{5}@{}}
    \toprule
        \multicolumn{2}{l}{Attendance Rate} &   & \multicolumn{2}{c}{Age 6-11} &   & \multicolumn{2}{c}{Age 12-15} &   & 					\multicolumn{2}{c}{Age 12-15, behind} &   & \multicolumn{2}{c}{Age 13-15, behind}  \\
        &  &  &  &  &  &  &  &  &  &  &  &  \multicolumn{2}{c}{HGC* $\geq$ 6}\\
    \cmidrule{4-5}
    \cmidrule{7-8}
    \cmidrule{10-11}
    \cmidrule{13-14}
        &  &  & \multicolumn{1}{c}{Girls} & \multicolumn{1}{c}{Boys} & & \multicolumn{1}{c}{Girls} & 
        \multicolumn{1}{c}{Boys} & & \multicolumn{1}{c}{Girls} & \multicolumn{1}{c}{Boys} & & 
        \multicolumn{1}{c}{Girls} & \multicolumn{1}{c}{Boys} \\
    \midrule
    \multicolumn{1}{l}{\textit{A. TW2006}} &  &  &  &  &  &  &  &  &  &  &  &\\
                             & Actual       &   & 96.5  & 96.7  &   & 66.5  & 72.5  &   & 58.7  & 67.4  &   & 44.4  & 57.1 \\
                             & Predicted    &   & 96.2  & 96.4  &   & 61.8  & 68.8  &   & 55.5  & 65.3  &   & 45.3  & 53.0  \\
                             & Err          &   & -0.3  & -0.3  &   & -4.7  & -3.7  &   & -3.2  & -2.1  &   &  0.9  & -4.1  \\
                                            &  &  &  &  &  &  &  &  &  &  &  &\\
	\multicolumn{1}{l}{\textit{B. Current}} &  &  &  &  &  &  &  &  &  &  &  &\\                             
							 & Actual       &   & 97.7  & 96.9  &   & 67.3  & 76.9  &   & 57.3  & 70.0  &   & 41.1  & 60.0 \\
    \multicolumn{1}{l}{CART} &              &   &       &       &   &       &       &   &       &       &   &       &      \\
                             & Predicted    &   & 100   & 100   &   & 86.5  & 81.2  &   & 70.0  & 70.2  &   & 73.3  & 30.4 \\
                             & Err          &   & 2.3   & 3.1   &   & 19.2  &  4.3  &   & 12.7  &  0.2  &   & 32.2  &-29.6 \\
    \multicolumn{1}{l}{C4.5} &              &   &       &       &   &       &       &   &       &       &   &       &     \\
                             & Predicted    &   & 98.4  & 94.4  &   & 75.4  & 81.4  &   & 58.6  & 70.1  &   & 53.6  & 65.2 \\
                             & Err          &   &  0.7  & -2.5  &   &  8.1  &  4.5  &   &  1.3  &  0.1  &   & 12.5  &  5.2 \\
    \multicolumn{1}{l}{LASSO}&              &   &       &       &   &       &       &   &       &       &   &       &     \\
                             & Predicted    &   & 93.4  & 95.7  &   & 82.4  & 81.6  &   & 61.3  & 65.3  &   & 71.0  & 50.0 \\
                             & Err          &   & -4.3  & -1.2  &   & 15.1  &  4.7  &   &  4.0  & -4.7  &   & 29.9  &-10.0\\
    \multicolumn{1}{l}{Random Forest}       &  &  &  &  &  &  &  &  &  &  &  &  &     \\
                             & Predicted    &   & 99.7  & 99.5  &   & 85.4  & 87.1  &   & 66.0  & 71.1  &   & 53.3  & 42.9 \\
                             & Err          &   &  2.0  &  2.6  &   & 18.1  & 10.2  &   &  8.7  &  1.1  &   & 12.2  &-17.1 \\
    \multicolumn{1}{l}{AdaBoost}&           &   &       &       &   &       &       &   &       &       &   &       &     \\
                             & Predicted    &   & 99.7  & 99.5  &   & 85.4  & 84.9  &   & 60.0  & 66.7  &   & 53.3  & 43.5 \\
                             & Err          &   &  2.0  &  2.6  &   & 18.1  &  8.0  &   &  2.7  & -3.3  &   & 12.2  &-16.5  \\
    \bottomrule
    \end{tabular}%
   \label{tab:NStep98control}
\end{sidewaystable}%

\begin{sidewaystable}[htbp]
  \centering
  \caption{$N$-Step-Ahead Simulation, 1997 Treatment Group}
    \begin{tabular}{@{}lll.{1}.{1}c.{1}.{1}c.{1}.{3}c.{1}.{5}@{}}
    \toprule
        \multicolumn{2}{l}{Attendance Rate} &   & \multicolumn{2}{c}{Age 6-11} &   & \multicolumn{2}{c}{Age 12-15} &   & 					\multicolumn{2}{c}{Age 12-15, behind} &   & \multicolumn{2}{c}{Age 13-15, behind}  \\
        &  &  &  &  &  &  &  &  &  &  &  &  \multicolumn{2}{c}{HGC* $\geq$ 6}\\
    \cmidrule{4-5}
    \cmidrule{7-8}
    \cmidrule{10-11}
    \cmidrule{13-14}
        &  &  & \multicolumn{1}{c}{Girls} & \multicolumn{1}{c}{Boys} & & \multicolumn{1}{c}{Girls} & 
        \multicolumn{1}{c}{Boys} & & \multicolumn{1}{c}{Girls} & \multicolumn{1}{c}{Boys} & & 
        \multicolumn{1}{c}{Girls} & \multicolumn{1}{c}{Boys} \\
    \midrule
    \multicolumn{1}{l}{\textit{A. TW2006}} &  &  &  &  &  &  &  &  &  &  &  &\\
                             & Actual       &   & 97.6  & 97.6  &   & 62.9  & 69.5  &   & 56.9  & 64.2  &   & 30.3  & 52.6 \\
                             & Predicted    &   & 96.4  & 96.3  &   & 61.8  & 68.0  &   & 55.6  & 62.7  &   & 37.3  & 51.7  \\
                             & Err          &   & -1.2  & -1.3  &   & -1.1  & -1.5  &   & -1.3  & -1.5  &   &  7.0  & -0.9  \\
                                            &  &  &  &  &  &  &  &  &  &  &  &\\
	\multicolumn{1}{l}{\textit{B. Current}} &  &  &  &  &  &  &  &  &  &  &  &\\                             
							 & Actual       &   & 98.7  & 98.3  &   & 71.3  & 73.2  &   & 62.6  & 64.0  &   & 41.1  & 50.4 \\
    \multicolumn{1}{l}{CART} &              &   &       &       &   &       &       &   &       &       &   &       &      \\
                             & Predicted    &   & 100   & 100   &   & 85.8  & 79.9  &   & 64.1  & 68.8  &   & 79.2  & 30.0 \\
                             & Err          &   & 1.3   & 1.7   &   & 14.5  &  6.7  &   &  1.5  &  4.8  &   & 38.1  &-20.4\\
    \multicolumn{1}{l}{C4.5} &              &   &       &       &   &       &       &   &       &       &   &       &     \\
                             & Predicted    &   & 97.6  & 94.7  &   & 70.3  & 76.0  &   & 40.9  & 57.9  &   & 33.3  & 53.3 \\
                             & Err          &   & -1.1  & -3.6  &   & -1.0  &  2.8  &   &-21.7  & -6.1  &   & -7.8   & 2.9 \\
    \multicolumn{1}{l}{LASSO}&              &   &       &       &   &       &       &   &       &       &   &       &     \\
                             & Predicted    &   & 96.0  & 92.5  &   & 83.0  & 75.8  &   & 60.0  & 56.3  &   & 69.8  & 50.8\\
                             & Err          &   & -2.7  & -5.8  &   & 11.7  &  2.6  &   & -2.6  & -7.7  &   & 28.7  &  0.4 \\
    \multicolumn{1}{l}{Random Forest}       &  &  &  &  &  &  &  &  &  &  &  &  &     \\
                             & Predicted    &   & 99.8  & 99.1  &   & 85.2  & 85.4  &   & 55.1  & 67.4  &   & 62.5  & 43.6 \\
                             & Err          &   &  1.1  &  0.8  &   & 13.9  & 12.2  &   & -7.5  &  3.4  &   & 21.4  & -6.8 \\
    \multicolumn{1}{l}{AdaBoost}&           &   &       &       &   &       &       &   &       &       &   &       &     \\
                             & Predicted    &   & 99.8  & 99.5  &   & 86.0  & 85.9  &   & 56.4  & 68.1  &   & 66.7  & 43.6 \\
                             & Err          &   &  1.1  &  1.2  &   & 14.7  & 12.7  &   & -6.2  &  4.1  &   & 25.6  & -6.8 \\
    \bottomrule
    \end{tabular}%
   \label{tab:NStep97treat}
\end{sidewaystable}%




\section{Conclusion}
We have compared the out-of-sample forecast between major machine learning models and the structural econometric model using the Progresa social experiment in this paper. The prediction performance of the machine learning models, in terms of the MAE and RMSE, is better than that found in the previous structural model used by \citet{Todd2006}. This finding provides answers to our two questions. Empirically, the Lucas Critique may not be quantitatively significant (at least for CCT policy), so machine learning may still be a reasonable tool for prediction or counter-factual analysis. Second, economists could benefit from machine learning approaches when the goal of a study is to improve the predictive power. However, the quality of the machine learning models depends heavily on the size of the data and the number of covariates. If either the data or covariates are limited, the structural model can give a more sensible prediction than machine learning models can. 

The result of one-step-ahead prediction also shows that within-sample fitness might not be a good indicator of a model's predictive power. A model with a good within-sample fitness might be too specific to the current data set and hence lose its generalization ability. It is important to keep a fraction of the data from model building and use it to test the out-of-sample forecast when trying to implement a new policy.

Moreover, machine learning algorithms are time-efficient in terms of model building. Even the most complicated model, the random forest model, takes only approximately half an hour to build without any parallel computing. In the era of big data, the advantage of machine learning models is increasingly apparent.

\clearpage
\bibliographystyle{jfe}
\bibliography{c:/xtemp/reference_dependent_preference}

\clearpage

\appendix
\section{Models of One-Step-Ahead Prediction}
\label{app:varInOneStep}
School Attendance Prediction Model:
\begin{align*}
	f(X,C) =  f(& parInc, parIncPerChild, dist2sch, dist2city, fAge, mAge, fAgeWed, mAgeWed, \\
		        & childNum, girlNum, childNumLe3, childNumLe3Sq, childNum35, childNum611, \\
		        & childNumLe15, avgBehindYrs, avgBehindGirl avgBehindBoy, avgSch615, \\
		        & avgSch615Girl, avgSch615Boy, avgSch615TimesChildNum, girlNumHgcGeq10, \\
	        	& childNumSecSch, childNumSecSchTimesDist2sch, childNum1215NotBehind, \\
	        	& preg, pregOrPrePreg, mAgeFristBirth, pregFirstYr, preg2024, preg2529,\\
	     	    & preg3034, preg3539, preg4043, minChildAge, girlGt11SecSchRatio,\\
	     	    & boyGt11SecSchRatio, childAge, gender, age1215, hgcGeq6, behindYrs)
\end{align*}

\begin{longtable}{@{}ll@{}}
  \caption{Variable Definitions}\label{tab:schVarDef}\\
    \toprule
        Variable &   Definition\\
    \midrule
    \endfirsthead
    \toprule
        Variable &   Definition\\
    \midrule
    \endhead
    \midrule
	\endfoot
	\endlastfoot
	    Household Characteristics, $X$ \\
        parInc 							& Parents' income\\
        parIncPerChild					& Average parent income for children less than 16 years old\\
        dist2sch						& Distance to secondary school\\
        dist2city						& Distance to city\\
        fAge							& Father's age during current period\\
        mAge							& Mother's age during current period\\
        fAgeWed							& Father's age when getting married\\
        mAgeWed							& Mother's age when getting married\\
        childNum						& Number of children in the household\\
        girlNum							& Number of girls in the household\\
        childNumLe3						& Number of children under age 3\\
        childNumLe3Sq					& Square of childNumLe3\\
        childNum35						& Number of children aged 3 to 5 years\\
        childNum611						& Number of children aged 6 to 11 years\\       
        childNumLe15					& Number of children aged 15 years or less \\
        avgBehindYrs					& The average number of years behind for children \\
        avgBehindGirl 					& The average number of years behind, girls older than age 6\\
        avgBehindBoy					& The average number of years behind, boys older than age 6\\
        avgSch615						& The average schooling years for children aged 6 to 15\\
        avgSch615Girl					& The average schooling years for girls aged 6 to 15\\
        avgSch615Boy					& The average schooling years for boys aged 6 to 15\\
        avgSch615TimesChildNum			& AvgSch615 $\times$ childNum\\
        girlNumHgcGeq10					& Number of girls with more than secondary school education\\
        childNumSecSch					& Number of children with more than primary school education\\
        childNumSecSchTimesDist2sch		& childNumSecSch $\times$ dist2sch\\
        childNum1215NotBehind			& Number of children who are 12 to 15 years old \\
        								& and not behind in school\\
        preg							& Pregnancy during current period\\
        pregOrPrePreg					& Pregnancy during current or last period\\
        mAgeFristBirth					& Mother's age when first giving birth\\
        pregFirstYr						& Indicator that a mother gives birth within \\
        								& the first year of marriage\\
        preg2024						& Pregnancy when a mother is between ages 20 and 24\\
        preg2529						& Pregnancy when a mother is between ages 25 and 29\\
        preg3034						& Pregnancy when a mother is between ages 30 and 34\\
        preg3539						& Pregnancy when a mother is between ages 35 and 39\\
        preg4043						& Pregnancy when a mother is between ages 40 and 43\\
        minChildAge						& The age of youngest children born to the household\\
        girlGt11SecSchRatio				& The ratio of girls who are older than 11 and \\
        								& ever attend secondary school\\
        boyGt11SecSchRatio				& The ratio of boys who are older than 11 and \\
        								& ever attend secondary school\\
        								& \\
        Personal Characteristics, $C$ \\
        childAge						& Age of child\\
        gender							& Gender of child\\
        age1215							& Indicator of child age between 12 to 15 years old \\
        hgcGeq6							& Indicator of the highest completed grade $\geq$ 6\\
        behindYrs						& The number of years behind for a child\\
    \bottomrule
\end{longtable}

\clearpage
\section{Models of $N$-Step-Ahead Prediction}
\label{app:varInNstep}
Income Prediction Model:
\begin{align*}
	f(X) = f(& fAge, fAgeSq, dist2sch, dist2city, dist2citySq, fAgeWed, mAgeWed, \\
		     & fAgeTimesDist2city, hgcParGe9, fAgeTimesDist2sch)
\end{align*}               

Pregnancy Prediction Model:
\begin{align*}
	f(X,C) = f(&parInc, parIncSq, mAge, mAgeSq, mAgeCube, fAge, fAgeSq, fAgeCube,\\
               &fAgeWed, mAgeWed, prePreg, minChildAge, dist2sch, dist2city, \\
               &childNum, childNumLe3, childNum35, childNum611, \\
               &parIncPerChild, yrAfterWed)
\end{align*}               

School Failure Prediction Model
\begin{align*}
	f(X,C) = f(&hgc, hgcSq, childAge, childAgeSq, gender, \\
			   &parInc, parIncSq, dis2city, dist2sch, hgcParGe9, \\
			   &age815Times0Sch, ageGe7, ageGe7TimesGender)
)
\end{align*}               

\begin{table}[htbp]
  \caption{Variable Definitions}\label{tab:incVarDef}
 \begin{tabularx}{\textwidth}{XX}
    \toprule
        Variable &   Definition\\
    \midrule
	    Household Characteristics, $X$ \\
	    hgcParGe9                       & Indicator of parents' education level higher than\\
	    								& secondary school\\
        fAgeTimesDist2city 				& fAge $\times$ dist2city\\
        fAgeTimesDist2sch 				& fAge $\times$ dist2sch\\
        yrAfterWed						& Years after getting married\\
                								& \\
        Personal Characteristics, $C$ \\
        age815Times0Sch					& Indicator of age 8 to 15 and not yet finished a grade\\
        ageGe7							& Indicator of age larger than 7\\
        ageGe7TimesGender				& ageGe7 $\times$ gender\\
    \bottomrule
  \end{tabularx}%
\end{table}

\begin{sidewaystable}[htbp]
  \centering
  \caption{Income Prediction Across Different Subgroups}
    \begin{tabular}{@{}lllrrcrrcrrcrr@{}}
    \toprule
        \multicolumn{2}{l}{Attendance Rate} &   & \multicolumn{2}{c}{Age 6-11} &   & \multicolumn{2}{c}{Age 12-15} &   & 					\multicolumn{2}{c}{Age 12-15, behind} &   & \multicolumn{2}{c}{Age 13-15, behind}  \\
        &  &  &  &  &  &  &  &  &  &  &  &  \multicolumn{2}{c}{HGC* $\geq$ 6}\\
    \cmidrule{4-5}
    \cmidrule{7-8}
    \cmidrule{10-11}
    \cmidrule{13-14}
        &  &  & \multicolumn{1}{c}{Girls} & \multicolumn{1}{c}{Boys} & & \multicolumn{1}{c}{Girls} & 
        \multicolumn{1}{c}{Boys} & & \multicolumn{1}{c}{Girls} & \multicolumn{1}{r}{Boys} & & 
        \multicolumn{1}{c}{Girls} & \multicolumn{1}{r}{Boys} \\
    \midrule
    \multicolumn{1}{l}{\textit{TW2006}} &  &  &  &  &  &  &  &  &  &  &  &\\
    \multicolumn{1}{l}{\mbox{\quad 1997 control}}      
	     & Actual    &   & 13774 & 14149 &   & 12146 & 12712 &   & 11999 & 11819 &   & 12246 & 12577 \\
         & Predicted &   & 13455 & 13433 &   & 13707 & 13783 &   & 13278 & 13417 &   & 13545 & 13881  \\
         & MAE       &   &  8666 &  9150 &   &  7477 &  7892 &   &  7245 &  7287 &   &  7740 &  7845  \\
         & RMSE      &   & 17677 & 17925 &   & 13478 & 14182 &   & 13896 & 12498 &   & 14807 & 14959  \\
    \multicolumn{1}{l}{\mbox{\quad 1998 control}}                                   
	     & Actual    &   & 11332 & 11356 &   & 11087 & 11888 &   & 10557 & 11169 &   & 11662 & 11315 \\
         & Predicted &   & 13504 & 13526 &   & 13862 & 13805 &   & 13541 & 13396 &   & 13699 & 13804  \\
         & MAE       &   &  6418 &  6408 &   &  6206 &  7445 &   &  6170 &  6992 &   &  6639 &  6481  \\
         & RMSE      &   & 11260 & 10056 &   &  8849 & 12178 &   &  9431 & 11136 &   & 11300 &  9044  \\
    \multicolumn{1}{l}{\mbox{\quad 1997 treatment}}                                                                
	     & Actual    &   & 11471 & 11027 &   & 12190 & 12793 &   & 11346 & 10978 &   & 11876 & 12179 \\
         & Predicted &   & 13257 & 13149 &   & 13483 & 13569 &   & 13112 & 13163 &   & 13415 & 13535  \\
         & MAE       &   &  6834 &  6584 &   &  7393 &  7899 &   &  6956 &  6742 &   &  7303 &  6950  \\
         & RMSE      &   & 10982 & 11247 &   & 11654 & 15382 &   & 10399 & 10490 &   & 11056 & 12086  \\
                             
                                            &  &  &  &  &  &  &  &  &  &  &  &\\
	\multicolumn{1}{l}{\textit{B. Current}} &  &  &  &  &  &  &  &  &  &  &  &\\                             
    \multicolumn{1}{l}{\mbox{\quad 1997 control}}      
	     & Actual    &   & 13774 & 14149 &   & 12146 & 12712 &   & 11999 & 11819 &   & 12246 & 12577 \\
         & Predicted &   & 12710 & 12580 &   & 12170 & 12476 &   & 11774 & 12097 &   & 11726 & 12623  \\
         & MAE       &   &  6103 &  6393 &   &  5253 &  5037 &   &  5127 &  4707 &   &  5285 &  4969  \\         
         & RMSE      &   & 13616 & 13939 &   & 10644 &  9969 &   & 10910 &  8606 &   & 11641 & 10285  \\
    \multicolumn{1}{l}{\mbox{\quad 1998 control}}                                   
	     & Actual    &   & 11332 & 11356 &   & 11087 & 11888 &   & 10557 & 11169 &   & 11662 & 11315 \\
         & Predicted &   & 12153 & 12560 &   & 12646 & 11991 &   & 12088 & 11681 &   & 12449 & 11684  \\
         & MAE       &   &  4108 &  4445 &   &  4086 &  4254 &   &  3801 &  3955 &   &  4218 &  3867  \\                  
         & RMSE      &   &  8255 &  7499 &   &  6241 &  8350 &   &  6429 &  7520 &   &  7671 &  6573  \\
    \multicolumn{1}{l}{\mbox{\quad 1997 treatment}}                                                                
	     & Actual    &   & 11471 & 11027 &   & 12190 & 12793 &   & 11346 & 10978 &   & 11876 & 12179 \\
         & Predicted &   & 11540 & 11425 &   & 11555 & 11913 &   & 11365 & 11156 &   & 11808 & 11558  \\
         & MAE       &   &  4429 &  4050 &   &  4811 &  5083 &   &  4561 &  4370 &   &  4854 &  4651  \\                           
         & RMSE      &   &  7755 &  7799 &   &  8582 & 10673 &   &  7774 &  7692 &   &  8312 &  8892  \\
    \bottomrule
    \end{tabular}%
   \label{tab:incErr}
\end{sidewaystable}%

\begin{longtable}{@{}lrrrrr@{}}
  \caption{Coefficients of Pregnancy Model}\label{tab:logitPregCoef}\\
    \toprule
        Pr(preg=1) &   Estimate & Std Err & z Value & Pr($>|z|$) &\\
    \midrule
    \endfirsthead
    \toprule
        Pr(preg=1) &   Estimate & Std Err & z Value & Pr($>|z|$) &\\
    \midrule
    \endhead
    \midrule
	\endfoot
	\endlastfoot
	(Intercept)   &  1.338e+01 & 3.541e+00 &  3.779 & 0.000157& *** \\
	parInc        &  -5.216e-05&  2.259e-05&  -2.309& 0.020919& *  \\
	parIncSq      &   4.511e-10& 1.349e-10 &  3.345 &0.000822 &***\\
	mAge          &  -1.186e+00&  3.849e-01&  -3.081& 0.002061& ** \\
	mAgeSq        &   5.178e-02&  1.371e-02&   3.778& 0.000158& ***\\
	mAgeCube      &  -6.371e-04&  1.576e-04&  -4.043& 5.28e-05& ***\\
	fAge          &  -1.925e-01&  1.614e-01&  -1.193& 0.233044&    \\
	fAgeSq        &  -7.784e-04&  3.940e-03&  -0.198& 0.843369&    \\
	fAgeCube      &   8.520e-06&  3.188e-05&   0.267& 0.789284&    \\
	fAgeWed       &   2.205e-01&  4.562e-02&   4.834& 1.34e-06& ***\\
	mAgeWed       &  -2.264e-01&  4.880e-02&  -4.639& 3.51e-06& ***\\
	hgcParGe9     &  -3.842e-01&  1.391e-01&  -2.762& 0.005747& ** \\
	prePreg       & -7.028e-01 & 1.521e-01 & -4.621 &3.81e-06 &***\\
	minChildAge   &  5.704e-02 & 1.021e-02 &  5.589 &2.29e-08 &***\\
	dist2sch      &   4.097e-05&  2.818e-05&   1.454& 0.145968&    \\
	dist2city     &  -1.278e-03&  7.503e-04&  -1.704& 0.088463& .  \\
	childNum      &  -2.105e+00&  2.937e-01&  -7.169& 7.56e-13& ***\\
	childNumSq    &   1.568e-01&  2.661e-02&   5.893& 3.79e-09& ***\\
	childNumLe3   &   7.058e+00&  4.463e-01&  15.814&  $<$ 2e-16& ***\\
	childNumLe3Sq &  -1.754e-01&  9.758e-02&  -1.797& 0.072289& .  \\
	childNum35    &   5.817e+00&  3.843e-01&  15.135&  $<$ 2e-16& ***\\
	childNum611   &   9.414e-01&  1.945e-01&   4.840& 1.30e-06& ***\\
	boyNum        &  -8.563e-02&  1.184e-01&  -0.723& 0.469388&    \\
	girlNum811Home&   6.701e+00&  6.381e-01&  10.502&  $<$ 2e-16& ***\\
	girlNum1215Home&  5.987e+00&  4.907e-01&  12.202&  $<$ 2e-16& ***\\
	childNumHome  &  -5.475e+00&  3.395e-01& -16.127&  $<$ 2e-16& ***\\
	girlNumHome   &  -6.679e-02&  1.537e-01&  -0.434& 0.663967&    \\
	parIncPerChild&  -4.792e-05&  2.779e-05&  -1.724& 0.084624& .  \\
    \bottomrule
\end{longtable}

\begin{table}[htbp]
\begin{threeparttable}
  \centering
  \caption{Pregnancy Prediction by Age}
    \begin{tabularx}{\textwidth}{@{}XX.{1}X.{1}.{1}@{}}
    \toprule
        &       & \multicolumn{1}{c}{Actual Pregnancy Rate} &
        &\multicolumn{2}{c}{Logit}\\
		\cmidrule{5-6}
		&       &        &         
		& \multicolumn{1}{c}{Predict} &  \multicolumn{1}{c}{Error} \\
    \midrule
  	\multicolumn{1}{l}{\textit{1997 control}} &  &  &  &  &  \\                             
    & Age 20-24  &  19.2 &   & 21.2  &  2.0 \\  
    & Age 25-29  &  15.3 &   & 12.1  & -3.2 \\
    & Age 30-34  &  12.1 &   &  6.5  & -5.6 \\
	& Age 35-44  &   4.9 &   &  4.5  & -0.4  \\
  	\multicolumn{1}{l}{\textit{1998 control}} &  &  &  &  &  \\                                      
	& Age 20-24  & 16.2  &   & 14.6  & -1.6 \\
	& Age 25-29  & 16.7  &   &  8.3  & -8.4  \\
	& Age 30-34  &  9.3  &   &  9.7  &  0.4  \\
	& Age 35-44  &  5.1  &   &  3.6  & -1.5   \\
  	\multicolumn{1}{l}{\textit{1997 treatment}} &  &  &  &  &  \\                                      
	& Age 20-24  & 18.7  &   & 17.0  & -1.7 \\
	& Age 25-29  & 17.2  &   & 13.1  & -4.1 \\
	& Age 30-34  & 11.3  &   &  9.3  & -2.0 \\
	& Age 35-44  &  4.9  &   &  5.1  &  0.2 \\
    \bottomrule
    \end{tabularx}%
  \label{tab:pregErr}%
\end{threeparttable}
\end{table}%

\begin{sidewaystable}[htbp]
  \centering
  \caption{School Failure Prediction Across Different Subgroups}
    \begin{tabular}{@{}lll.{1}.{1}c.{1}.{1}c.{1}.{3}c.{1}.{5}@{}}
    \toprule
        \multicolumn{2}{l}{Attendance Rate} &   & \multicolumn{2}{c}{Age 6-11} &   & \multicolumn{2}{c}{Age 12-15} &   & 					\multicolumn{2}{c}{Age 12-15, behind} &   & \multicolumn{2}{c}{Age 13-15, behind}  \\
        &  &  &  &  &  &  &  &  &  &  &  &  \multicolumn{2}{c}{HGC* $\geq$ 6}\\
    \cmidrule{4-5}
    \cmidrule{7-8}
    \cmidrule{10-11}
    \cmidrule{13-14}
        &  &  & \multicolumn{1}{c}{Girls} & \multicolumn{1}{c}{Boys} & & \multicolumn{1}{c}{Girls} & 
        \multicolumn{1}{c}{Boys} & & \multicolumn{1}{c}{Girls} & \multicolumn{1}{c}{Boys} & & 
        \multicolumn{1}{c}{Girls} & \multicolumn{1}{c}{Boys} \\
    \midrule
    \multicolumn{1}{l}{\textit{TW2006}} &  &  &  &  &  &  &  &  &  &  &  &\\
    \multicolumn{1}{l}{\mbox{\quad 1997 control}}      
	     & Actual    &   & 15.2 & 18.3 &   & 20.1 & 14.8 &   & 20.7 & 13.0 &   & 30.6 &  9.4 \\
         & Predicted &   & 13.4 & 15.1 &   & 13.6 & 13.6 &   & 14.0 & 14.9 &   & 13.4 & 13.1  \\
         & Err       &   & -1.8 & -3.2 &   & -6.5 & -1.2 &   & -6.7 &  1.9 &   &-17.2 &  3.7  \\
    \multicolumn{1}{l}{\mbox{\quad 1998 control}}                                   
	     & Actual    &   & 11.8 & 14.8 &   & 15.1 & 10.0 &   & 16.0 &  7.8 &   & 20.0 &  2.6 \\
         & Predicted &   & 12.9 & 14.7 &   & 13.4 & 13.5 &   & 13.6 & 15.3 &   & 13.6 & 13.1  \\
         & Err       &   & 	1.1 & -0.1 &   & -1.7 &  3.5 &   & -2.4 &  7.5 &   & -6.4 & 10.5  \\
    \multicolumn{1}{l}{\mbox{\quad 1997 treatment}}                                                                
	     & Actual    &   & 12.8 & 15.7 &   & 15.8 &  9.5 &   & 18.8 &  6.3 &   & 21.6 &  9.7 \\
         & Predicted &   & 13.2 & 14.7 &   & 14.6 & 13.9 &   & 16.5 & 16.1 &   & 14.8 & 12.9  \\
         & Err       &   &  0.4 & -1.0 &   & -1.2 &  4.4 &   & -2.3 &  9.8 &   & -6.8 &  3.2  \\
                                            &  &  &  &  &  &  &  &  &  &  &  &\\
	\multicolumn{1}{l}{\textit{B. Current}} &  &  &  &  &  &  &  &  &  &  &  &\\                             
    \multicolumn{1}{l}{\mbox{\quad 1997 control}}      
	     & Actual    &   & 15.2 & 18.3 &   & 20.1 & 14.8 &   & 20.7 & 13.0 &   & 30.6 &  9.4 \\
         & Predicted &   & 11.3 & 18.2 &   & 19.0 &  9.4 &   & 25.0 &  8.1 &   & 28.9 & 11.3  \\
         & Err       &   & -3.9 & -0.1 &   & -1.1 & -5.4 &   &  4.3 & -4.9 &   & -1.7 &  1.9  \\
    \multicolumn{1}{l}{\mbox{\quad 1998 control}}                                   
	     & Actual    &   & 11.8 & 14.8 &   & 15.1 & 10.0 &   & 16.0 &  7.8 &   & 20.0 &  2.6 \\
         & Predicted &   &  9.4 & 14.6 &   & 17.1 &  7.6 &   & 17.3 &  6.8 &   & 23.3 &  5.1  \\
         & Err       &   & -2.4 & -0.2 &   &  2.0 & -2.4 &   &  1.3 & -1.0 &   &  3.3 &  2.5  \\
    \multicolumn{1}{l}{\mbox{\quad 1997 treatment}}                                                                
	     & Actual    &   & 12.8 & 15.7 &   & 15.8 &  9.5 &   & 18.8 &  6.3 &   & 21.6 &  9.7 \\
         & Predicted &   &  8.7 & 16.6 &   & 15.4 &  6.0 &   & 20.5 &  3.1 &   & 32.4 &  3.2  \\
         & Err       &   & -4.1 &  0.9 &   & -0.4 & -3.5 &   &  1.7 & -3.2 &   & 10.8 & -6.5  \\
    \bottomrule
    \end{tabular}%
   \label{tab:failErr}
\end{sidewaystable}%

\clearpage
\section{Machine Learning Models for Out-of-Sample Forecasts}
\label{app:ML_model}
\begin{figure}[htbp]
    \centering
    \includegraphics[width=1.0\linewidth]{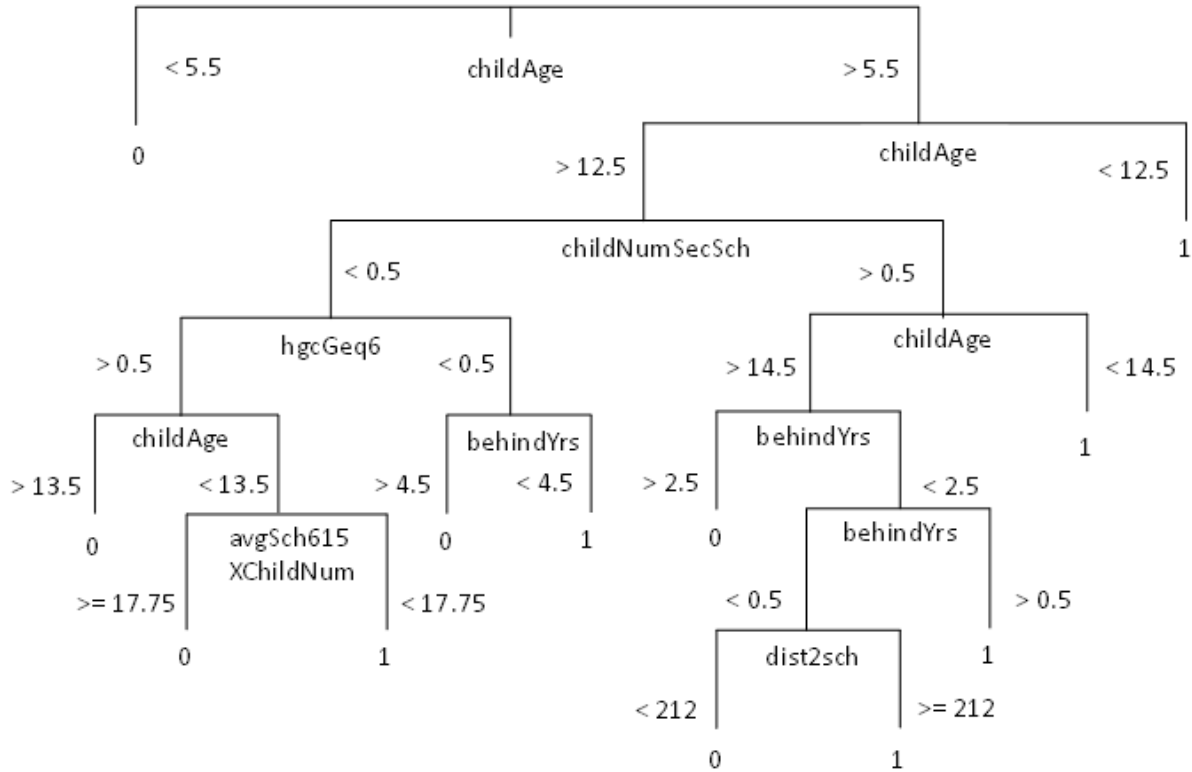}
    \caption{CART Model} \label{fig:CART_train}
\end{figure}

\begin{longtable}{@{}ll@{}}
  \caption{Coefficients of LASSO Model}\label{tab:lassoCoef}\\
    \toprule
        Variable &   Estimate\\
    \midrule
    \endfirsthead
    \toprule
        Variable &   Estimate\\
    \midrule
    \endhead
    \midrule
	\endfoot
	\endlastfoot
	    Household Characteristics, $X$ \\
        parInc 							& 1.2472e-6\\
        dist2sch						& -1.9364e-5\\
        dist2city						& 1.5016e-3\\
        fAge							& -6.0436e-3\\
        mAge							& -3.3402e-2\\
        mAgeWed							& 2.2619e-2\\
        childNum35						& -0.8414\\
        childNum611						& 0.2936 \\       
        childNumLe15					& -4.1092e-3\\
        avgBehindYrs					& 0.1571 \\
        avgBehindGirl 					& -1.3050e-3\\
        avgSch615TimesChildNum			& -9.4094e-3\\
        girlNumHgcGeq10					& 0.2257\\
        childNumSecSch					& 0.1617\\
        childNum1215NotBehind			& 0.1269\\
        preg2529						& -8.5063e-3\\
        girlGt11SecSchRatio				& -1.048e-2\\
        childAge						& 1.5737\\
        age1215							& -4.4320\\
        hgcGeq6							& -5.4719\\
        behindYrs						& -1.9175\\
    \bottomrule
\end{longtable}

\begin{figure}[htbp]
    \centering
    \includegraphics[width=1.0\linewidth]{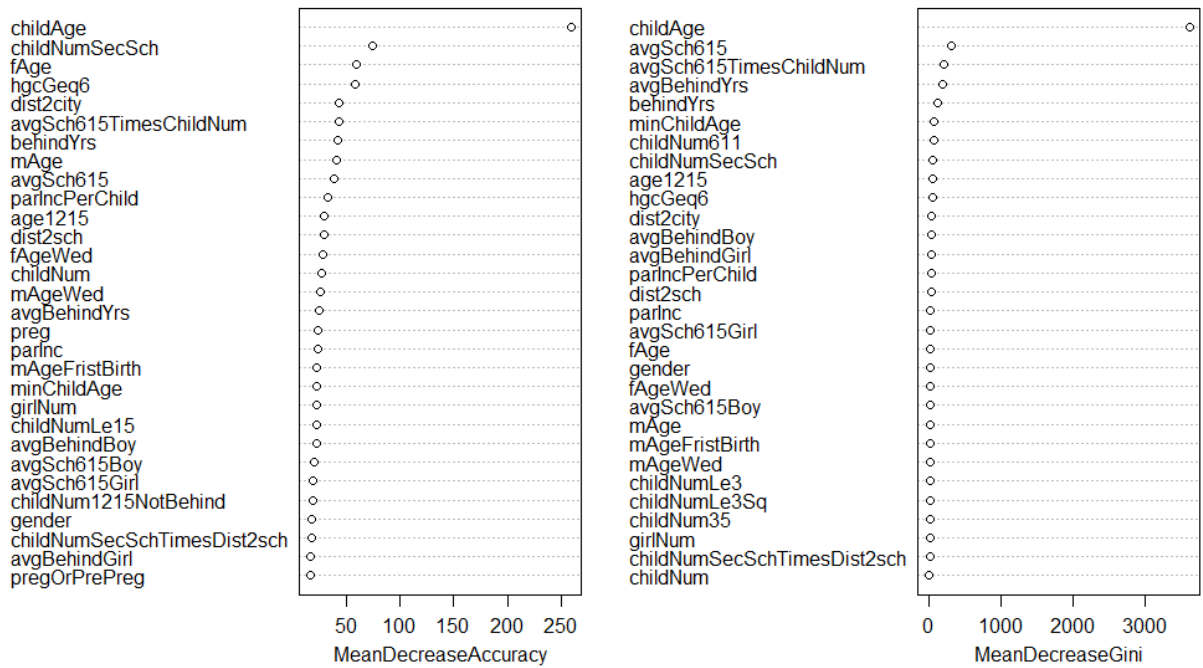}
    \caption{Variable Importance for Random Forest Model}
    \vspace{2pt}
    \begin{minipage}{\linewidth}
    \footnotesize
    \emph{The importance of variables is sorted in descending order, 
    	according to either the mean decrease in accuracy (left) 
    	or the mean decrease in Gini gain (right) of the out-of-bag samples.}
    \end{minipage}
    \label{fig:rfVar}
\end{figure}

\end{document}